\documentclass[12pt,a4paper,reqno]{amsart}
\usepackage[utf8]{inputenc}

\usepackage{amsmath,amssymb,a4wide,amsthm,ifthen,hyperref,enumerate,enumitem,caption}

\usepackage{epstopdf}% To incorporate .eps illustrations using PDFLaTeX, etc.
\usepackage[section]{placeins}

\usepackage[numbers,sort&compress]{natbib}% Citation support using natbib.sty
\bibpunct[, ]{[}{]}{,}{n}{,}{,}% Citation support using natbib.sty
\makeatletter% @ becomes a letter
\def\NAT@def@citea{\def\@citea{\NAT@separator}}% Suppress spaces between citations using natbib.sty
\makeatother% @ becomes a symbol again

\theoremstyle{plain}% Theorem-like structures provided by amsthm.sty

\theoremstyle{definition}

\theoremstyle{remark}

\usepackage{graphicx}
\usepackage{url}
\usepackage{ulem}
\usepackage{epstopdf}
\usepackage[section]{placeins}
\usepackage{natbib}
\usepackage{flafter} 
\usepackage{multirow}
\usepackage{booktabs}
\usepackage[table,xcdraw]{xcolor}

\begin{document}

\title{Variable selection in a specific regression time series of counts}

\author{M. Gomtsyan} 
\address{Université Paris-Saclay, AgroParisTech, INRAE, UMR MIA Paris-Saclay, 91120 Palaiseau, France}

\begin{abstract}
Time series of counts occurring in various applications are often overdispersed, meaning their variance is much larger than the mean. This paper proposes a novel variable selection approach for processing such data. Our approach consists in modelling them using sparse negative binomial GLARMA models. It combines estimating the autoregressive moving average (ARMA) coefficients of GLARMA models and the overdispersion parameter with performing variable selection in regression coefficients of Generalized Linear Models (GLM) with regularised methods. We describe our three-step estimation procedure, which is implemented in the \texttt{NBtsVarSel} package. We evaluate the performance of the approach on synthetic data and compare it to other methods. Additionally, we apply our approach to RNA sequencing data. Our approach is computationally efficient and outperforms other methods in selecting variables, i.e. recovering the non-null regression coefficients. 
\end{abstract}

\keywords{count time series, negative binomial distribution, variable selection}

\maketitle

\section{Introduction}

In recent years, the interest in the study of count time series has increased. These series represent a record of the number of occurrences of events over time and, consequently, are nonnegative and integer-valued. They find practical applications in various fields, such as the contagion dynamics of COVID-19 in epidemiology~\cite{agosto:giudici:2020}, the number of transactions in stocks in finance~\cite{brannas:quoreshi:2010}, and RNA sequencing (RNA-Seq) kinetics data in molecular biology~\cite{Thorne:2018}. 

Count time series require special treatment since many continuous models cannot interpret discrete data~\cite{handbook:2016}. In addition, as mentioned in~\cite{davis:etal:2021}, count time series are often overdispersed, i.e. the variance is larger than the mean. One can capture the overdispersed nature of such data with negative binomial distribution models. In particular, they efficiently interpret RNA-Seq data~\cite{love:huber:anders:2014, robinson:mccarthy:smyth:2009}. %Inspired by the study of RNA-Seq data, in this paper we propose a novel negative binomial count time series model.

Numerous models exist for count time series, with a detailed review in~\cite{davis:etal:2021}. These models can be grouped into two main classes: Integer Autoregressive Moving Average (INARMA) and generalized state space models. McKenzie in~\cite{McKenzie:1985} and Al-Osh and Alzaid in~\cite{al-osh:alzaid:1987} were the first to study the Integer Autoregressive process (INAR(1)). Later in~\cite{al-osh:alzaid:1990} it was extended to $p$th order process. The Integer-valued Moving Average (INMA) process is introduced in \cite{Al-Osh:1988}. Integer-valued generalized autoregressive conditional heteroskedasticity (INGARCH) models that can handle overdispersion are studied in~\cite{weis:2008} and~\cite{zhu:2011}. An advantage of INARMA processes is their autocorrelation structure, which is similar to the one of the autoregressive moving average (ARMA) models. However, the statistical inference in INAR models is more complex, as explained in~\cite{davis:etal:2021}. It requires intensive computational approaches, such as the efficient MCMC algorithm in~\cite{Neal:rao:2007}, developed for INARMA processes of known AR and MA orders. We refer the reader to~\cite{weiss:dts} for further details on INARMA models. 

Generalized state space models, introduced in~\cite{cox:1981}, are one of the most commonly used approaches for time series analysis~\cite{davis:etal:2021}. These models can be classified as parameter-driven and observation-driven models. The main difference between these two model groups is that the state vector evolves independently of past observations in parameter-driven models. In contrast, in observation-driven models, the state vector depends on the past history of the observations.

An overview of parameter-driven models can be found in~\cite{davis:1999}. In \cite{zeger:1988}, the Poisson log-liner regression is introduced, which in~\cite{blais:2000} is extended to the case where observations are assumed to have a distribution from the exponential family. In \cite{davis:2009}, Davis and Wu considered a negative binomial model, where the serial dependence is introduced through a dependent latent process in the link function. Despite the simple construction of these models, the parameter estimation in parameter-driven models is computationally expensive, as explained in~\cite{jung:2001}. 

The observation-driven models do not suffer from this computational downside. Following the introduction in~\cite{cox:1981}, they were further studied in~\cite{zeger:qaqish:1988}. In the literature, there are two types of observation-driven models: the Generalized Linear Autoregressive Moving Average (GLARMA) models introduced in~\cite{davis:1999} and further studied in~\cite{davis:dunsmuir:streett:2003}, \cite{davis:dunsmuir:street:2005}, \cite{dunsmuir:2015} and  the \mbox{(log-)linear} Poisson autoregressive models studied in~\cite{fokianos:2009}, \cite{fokianos:2011} and~\cite{fokianos:2012}. Note that GLARMA models cannot be seen as a particular case of the log-linear Poisson autoregressive models. 

In this paper, we will consider the negative binomial GLARMA model introduced in \cite{davis:dunsmuir:street:2005} with additional covariates. More precisely, given the past history $\mathcal{F}_{t-1}=\sigma(Y_s,s\leq t-1)$, we assume that
\begin{equation}\label{eq:Yt}
Y_t|\mathcal{F}_{t-1}\sim\text{NB}\left(\mu_t^\star, \alpha^\star\right),
\end{equation}
where $\text{NB}(\mu, \alpha)$ denotes the negative binomial distribution with mean $\mu$ and overdispersion parameter $\alpha$. In (\ref{eq:Yt}),
\begin{equation}\label{eq:mut_Wt}
\mu_t^\star=\exp(W_t^\star) \textrm{ with } W_t^\star=\sum_{i=0}^p\beta_i^\star x_{t,i}+Z_t^\star.
\end{equation}
Here the $x_{t,i}$'s represent the $p$ regressor variables ($p\geq 1$) and
\begin{equation}\label{eq:Zt}
Z_t^\star=\sum_{j=1}^q \gamma_j^\star E_{t-j}^\star \textrm{ with } E_t^\star=\frac{Y_t-\mu_t^\star}{\mu_t^\star + {\mu_t^\star}^2/\alpha^\star},
\end{equation}
where $1\leq q\leq\infty$ and $E_t^\star=0$ for all $t\leq 0$. The $E_t^\star$'s correspond to the working residuals in classical Generalized Linear Models (GLM). There are several types of residuals but in our model we consider score-type residuals, as proposed in \cite{koopman:2008}. It is important to mention that when $q=\infty$, $(Z_t^\star)$ satisfies the ARMA-like recursions provided in Equation (4) of \cite{davis:dunsmuir:street:2005}. The resulting model defined by (\ref{eq:Yt}), (\ref{eq:mut_Wt}) and (\ref{eq:Zt}) is the negative binomial GLARMA model.

The main goal of this paper is to introduce a novel approach for variable selection in the deterministic part (covariates) of sparse negative binomial GLARMA models defined in Equations \eqref{eq:Yt}, \eqref{eq:mut_Wt} and \eqref{eq:Zt}. Here the vector of the $\beta_i^\star$'s is sparse, i.e. many $\beta_i^\star$’s are null, and thus only a few regressor variables are explanatory. The novel approach that we propose consists in combining a procedure for estimating the ARMA part coefficients (to take into account the temporal dependence that may exist in the data) with regularised methods designed for GLM, as those proposed in \cite{friedman:hastie:tibshirani:2010} and \cite{hastie2019statistical}. The existing variable selection approaches for discrete data, such as~\cite{friedman:hastie:tibshirani:2010}, do not consider temporal dependence. 

Our approach can be applied in modelling RNA-Seq time series data in molecular biology. With RNA-Seq, it is possible to count the numbers of RNA fragments present in a biological sample. Linking these RNA fragments to genes allows  for determining the expression level of genes as integer counts. As explained in~\cite{wu2017diversity}, non-coding genes are potential key regulators of the expression of coding genes. In this framework, only a few among many non-coding genes are likely to be involved in explaining the expression of the coding genes. Since, as discussed earlier, the nature of RNA-Seq data is captured well with negative binomial models, a variable selection approach for sparse negative binomial GLARMA models can be efficient in identifying the relevant non-coding genes.

The paper is organised as follows. Firstly, in Section \ref{sec:estim}, we describe the properties of the likelihood of negative binomial GLARMA models. Secondly, in Section \ref{sec:our_estim} we propose a novel three-stage estimation procedure. It consists in first estimating the ARMA coefficients, then in estimating the regression coefficients by using a regularized approach, and estimating overdispersion parameter with a maximum likelihood approach. The algorithmic implementation of the methodology is given in Section \ref{subsec:algo}. Next, in Section \ref{sec:num}, we provide some numerical experiments on simulated data in order to illustrate our method and to compare its performance to the regularized methods designed for GLM of \cite{friedman:hastie:tibshirani:2010}. Finally, in Section \ref{sec:application}, we illustrate our method on RNA-Seq data that follows the temporal evolution of gene expression.

\section{Variable selection in sparse negative binomial GLARMA models}\label{sec:stat_inf}
In this section we introduce our variable selection approach in sparse negative binomial GLARMA models. We start by discussing the properties of the likelihood of negative binomial GLARMA models in Section~\ref{sec:estim}. Next, in Section~\ref{sec:our_estim} we explain how our approach estimates the parameters of the model. We  conclude by the description of the algorithm of our methodology in Section~\ref{subsec:algo}.

\subsection{Properties of the likelihood of negative binomial GLARMA models}\label{sec:estim}

As stated in \cite{glarma:package}, the probability mass function of negative binomial distribution is
  \begin{equation*}
  f(Y_t | W_t, \alpha) = \frac{\Gamma(\alpha + Y_t)}{\Gamma(\alpha) \Gamma(Y_t+1)} \Bigg(\frac{\alpha}{\alpha + \mu_t} \Bigg)^{\alpha} \Bigg(\frac{\mu_t}{\alpha + \mu_t} \Bigg)^{Y_t}.
  \end{equation*}
Note that it converges to the Poisson probability mass function when $\alpha \rightarrow \infty$.

Let us consider the parameter $\boldsymbol{\delta}^\star=(\boldsymbol{\beta}^{\star\prime},\boldsymbol{\gamma}^{\star\prime})$, where ${u}^\prime$ denotes the transpose of the vector $u$, $\boldsymbol{\beta}^\star=(\beta_0^\star,\beta_1^\star,\dots,\beta_p^\star)'$ represents the vector of regressor coefficients defined in (\ref{eq:mut_Wt}), and $\boldsymbol{\gamma}^\star=(\gamma_1^\star,\dots,\gamma_q^\star)'$ is the vector of the ARMA part coefficients defined in (\ref{eq:Zt}). Inspired by \cite{davis:dunsmuir:street:2005}, we will estimate $\boldsymbol{\delta}^\star$ by maximizing with respect to $\boldsymbol{\delta}=(\boldsymbol{\beta}',\boldsymbol{\gamma}')$, with $\boldsymbol{\beta}=(\beta_0,\beta_1,\dots,\beta_p)'$ and $\boldsymbol{\gamma}=(\gamma_1,\dots,\gamma_q)'$ the following criterion based on the conditional log-likelihood:
\begin{align}
\label{eq:likelihood}
L(\boldsymbol{\delta}, \alpha) 
&= \sum_{t=1}^n \Big( \log \Gamma(\alpha + Y_t) - \log\Gamma(Y_t+1) -\log\Gamma(\alpha)   \nonumber \\
&\quad \quad \quad + \alpha \log \alpha + Y_t W_t - (\alpha + Y_t) \log(\alpha + \exp(W_t)) \Big).
\end{align}
In (\ref{eq:likelihood}),
\begin{equation}\label{eq:Wt}
W_t(\boldsymbol{\delta}, \alpha)=\boldsymbol{\beta}'x_t+Z_t(\boldsymbol{\delta}, \alpha)=\beta_0+\sum_{i=1}^p\beta_i x_{t,i}+\sum_{j=1}^q \gamma_j E_{t-j}(\boldsymbol{\delta}, \alpha),
\end{equation}
with $x_t=(x_{t,0},x_{t,1},\dots,x_{t,p})'$, $x_{t,0}=1$ for all $t$ and
\begin{eqnarray}
E_t(\boldsymbol{\delta}, \alpha)=\frac{Y_t\exp(-W_t(\boldsymbol{\delta}, \alpha))-1}{1 + \frac{\exp(W_t(\pmb{\delta}, \alpha))}{\alpha}},\mbox{ if }t>0\mbox{ and }E_t(\boldsymbol{\delta}, \alpha)=0\mbox{, if }t\leq 0.
\label{eq:Et}
\end{eqnarray}

To obtain $\widehat{\boldsymbol{\delta}}$ defined by
\begin{equation*}
\widehat{\boldsymbol{\delta}}= \underset{\pmb{\delta}}{\arg\max} \; L(\boldsymbol{\delta}, \alpha),
\end{equation*}
we consider the first derivatives of $L$:
\begin{align}\label{eq:def:grad}
\frac{\partial L}{\partial \boldsymbol{\delta}}(\boldsymbol{\delta}, \alpha)& = \sum_{t=1}^n\left( Y_t \frac{\partial W_t(\boldsymbol{\delta}, \alpha)}{\partial \boldsymbol{\delta}} - \frac{(\alpha + Y_t) \exp(W_t(\boldsymbol{\delta}, \alpha))}{\alpha + \exp(W_t(\boldsymbol{\delta}, \alpha))} \frac{\partial W_t}{\partial \boldsymbol{\delta}}(\boldsymbol{\delta}, \alpha)\right) \nonumber \\
&= \sum_{t=1}^n\left( Y_t - \frac{(\alpha + Y_t) \exp(W_t(\boldsymbol{\delta}, \alpha))}{\alpha + \exp(W_t(\boldsymbol{\delta}, \alpha))} \right) \frac{\partial W_t}{\partial \boldsymbol{\delta}} ,
\end{align}
where 
\begin{equation*}
\frac{\partial W_t}{\partial \boldsymbol{\delta}}(\boldsymbol{\delta}, \alpha)=\frac{\partial\boldsymbol{\beta}' x_t}{\partial \boldsymbol{\delta}}+\frac{\partial Z_t}{\partial \boldsymbol{\delta}}
(\boldsymbol{\delta}, \alpha),
\end{equation*}
$\boldsymbol{\beta}$, $x_t$ and $Z_t$ being given in (\ref{eq:Wt}). The computations of the first derivatives of $W_t$ are detailed in Appendix \ref{subsub:first_derive}.

The Hessian of $L$ can be obtained as follows:
\begin{align}
\label{eq:def:hess}
\frac{\partial^2 L}{\partial \boldsymbol{\delta}'\partial \boldsymbol{\delta}}(\boldsymbol{\delta}, \alpha)
&= \sum_{t=1}^n \left( Y_t - \frac{(\alpha + Y_t) \exp(W_t(\boldsymbol{\delta}, \alpha))}{\alpha + \exp(W_t(\boldsymbol{\delta}, \alpha))} \right) \frac{\partial^2 W_t}{\partial \boldsymbol{\delta}'\partial\boldsymbol{\delta}}(\boldsymbol{\delta}, \alpha) \nonumber \\
&- \sum_{t=1}^n \frac{(\alpha + Y_t) \exp(W_t(\boldsymbol{\delta}, \alpha))}{\alpha + \exp(W_t(\boldsymbol{\delta}, \alpha))} \left( 1 - \frac{\exp(W_t(\boldsymbol{\delta}, \alpha))}{\alpha + \exp(W_t(\boldsymbol{\delta}, \alpha))}  \right)  \frac{\partial W_t}{\partial \boldsymbol{\delta}'}(\boldsymbol{\delta}, \alpha)\frac{\partial W_t}{\partial \boldsymbol{\delta}}(\boldsymbol{\delta}, \alpha).
\end{align}
The details for computing the second derivative of $W_t$ are given in Appendix \ref{subsub:second_derive}. 

Since in the sparse framework, with many components of $\boldsymbol{\beta}^\star$ being null, this procedure provides poor estimation results, we devised a novel estimation procedure described in the next section.

\subsection{Parameter estimation and variable selection}\label{sec:our_estim}

To select the most relevant elements of $\boldsymbol{\beta}^\star$, we propose a three-stage procedure. Firstly, we estimate $\boldsymbol{\gamma}^\star$ by using the Newton-Raphson algorithm described in Section \ref{sec:estim_gamma}. Next, we estimate $\boldsymbol{\beta}^\star$ by using the regularized approach outlined in Section \ref{sec:variable}. Finally, we estimate $\alpha^\star$ by a maximum likelihood approach as explained in Section \ref{sec:estim_alpha}. Additionally, in Section \ref{subsec:stab_sel} we explain how to guarantee the robustness of the selected variables.

\subsubsection{Estimation of $\boldsymbol{\gamma}^\star$}\label{sec:estim_gamma}

In order to obtain the estimate  of $\boldsymbol{\gamma}^\star$, we propose using
\begin{equation*}%\label{eq:delta_hat}
\widehat{\boldsymbol{\gamma}}=\underset{\pmb{\gamma}}{\arg\max} \; L({\boldsymbol{\beta}^{(0)}}',\boldsymbol{\gamma}', \alpha^{(0)}),
\end{equation*}
where $L$ is defined in (\ref{eq:likelihood}), $\boldsymbol{\beta}^{(0)}=(\beta_{0}^{(0)},\dots,\beta_{p}^{(0)})'$ and $\alpha^{(0)}$ are given initial estimations of $\boldsymbol{\beta}^\star$ and $\alpha^\star$, respectively, and $\boldsymbol{\gamma}=(\gamma_1,\dots,\gamma_q)'$. In Section \ref{subsec:algo} we explain how we choose these initial values. We use the Newton-Raphson algorithm to obtain $\widehat{\boldsymbol{\gamma}}$. For $r \geq 1$, starting from the initial value $\boldsymbol{\gamma}^{(0)}=(\gamma_1^{(0)},\dots,\gamma_q^{(0)})'$:
\begin{equation}
\label{eq:newton_raphson:gamma}
  \boldsymbol{\gamma}^{(r)}=\boldsymbol{\gamma}^{(r-1)}-\frac{\partial^2 L}{\partial \boldsymbol{\gamma}'\partial
    \boldsymbol{\gamma}}({\boldsymbol{\beta}^{(0)}}',{\boldsymbol{\gamma}^{(r-1)}}', \alpha^{(0)})^{-1}
  \frac{\partial L}{\partial \boldsymbol{\gamma}}({\boldsymbol{\beta}^{(0)}}',{\boldsymbol{\gamma}^{(r-1)}}', \alpha^{(0)}),
\end{equation}
where the first and second derivatives of $L$ are obtained using the same strategy as the one used
for deriving Equations (\ref{eq:def:grad}) and (\ref{eq:def:hess}) in Section \ref{sec:estim}.

\subsubsection{Variable selection: Estimation of $\boldsymbol{\beta}^\star$}\label{sec:variable} 

In order to obtain a sparse estimator of the $\beta_i^\star$'s in Model (\ref{eq:mut_Wt}), we use a regularized variable selection approach proposed in~\cite{friedman:hastie:tibshirani:2010} for fitting generalized linear models.  

To perform variable selection in the $\beta_i^\star$'s of Model (\ref{eq:mut_Wt}), in other words, to obtain a sparse estimator of $\pmb{\beta}^\star$, we shall use a methodology inspired by \cite{friedman:hastie:tibshirani:2010} for fitting generalized linear models. This approach penalises with $\ell_1$ penalties a quadratic approximation to the log-likelihood obtained by a Taylor expansion. Using $\boldsymbol{\beta}^{(0)}$, $\widehat{\boldsymbol{\gamma}}$, and $\alpha^{(0)}$ defined in Section \ref{sec:estim_gamma} the quadratic approximation is obtained as follows:
\begin{align*}
  \widetilde{L}(\boldsymbol{\beta})&:=L(\beta_0,\dots,\beta_p,\widehat{\gamma}, \alpha^{(0)})\\
  &=\widetilde{L}(\boldsymbol{\beta}^{(0)})
+\frac{\partial L}{\partial \boldsymbol{\beta}}(\boldsymbol{\beta}^{(0)},\widehat{\boldsymbol{\gamma}}, \alpha^{(0)})(\boldsymbol{\beta}-\boldsymbol{\beta}^{(0)})
+\frac12 (\boldsymbol{\beta}-\boldsymbol{\beta}^{(0)})'
\frac{\partial^2 L}{\partial \boldsymbol{\beta}\partial \boldsymbol{\beta}'}(\boldsymbol{\beta}^{(0)},\widehat{\boldsymbol{\gamma}}, \alpha^{(0)})
(\boldsymbol{\beta}-\boldsymbol{\beta}^{(0)}),
\end{align*}
where
$$\frac{\partial L}{\partial \boldsymbol{\beta}}=\left(\frac{\partial L}{\partial \beta_0},\dots,\frac{\partial L}{\partial \beta_p}\right)
\textrm{ and }
\frac{\partial^2 L}{\partial \boldsymbol{\beta}\partial \boldsymbol{\beta}'}=\left(\frac{\partial^2 L}{\partial \beta_j \partial \beta_k}\right)_{0\leq j,k\leq p}.$$
Hence we get,
\begin{align}\label{eq:Ltilde}
\widetilde{L}(\boldsymbol{\beta})=\widetilde{L}(\boldsymbol{\beta}^{(0)})+\frac{\partial L}{\partial \boldsymbol{\beta}}(\boldsymbol{\beta}^{(0)},\widehat{\boldsymbol{\gamma}}, \alpha^{(0)})
U(\boldsymbol{\nu}-\boldsymbol{\nu}^{(0)})-\frac12 (\boldsymbol{\nu}-\boldsymbol{\nu}^{(0)})' \Lambda (\boldsymbol{\nu}-\boldsymbol{\nu}^{(0)}),
\end{align}
where $U\Lambda U'$ is the singular value decomposition of the positive semidefinite symmetric matrix 
$-\frac{\partial^2 L}{\partial \boldsymbol{\beta}\partial \boldsymbol{\beta}'}(\boldsymbol{\beta}^{(0)},\widehat{\boldsymbol{\gamma}}, \alpha^{(0)})$
and $\boldsymbol{\nu}-\boldsymbol{\nu}^{(0)}=U'(\boldsymbol{\beta}-\boldsymbol{\beta}^{(0)})$.

In order to obtain a sparse estimator $\widehat{\pmb{\beta}}$ of $\boldsymbol{\beta}^\star$, we use the criterion $\widehat{\boldsymbol{\beta}}(\lambda)$ defined by
\begin{equation}\label{eq:beta_hat}
\widehat{\boldsymbol{\beta}}(\lambda)=\textrm{Argmin}_{\boldsymbol{\beta}}\left\{-\widetilde{L}_Q(\boldsymbol{\beta})+\lambda \|\boldsymbol{\beta}\|_1\right\},
\end{equation}
for a positive $\lambda$, where $\|\boldsymbol{\beta}\|_1=\sum_{k=0}^p |\beta_k|$ and $\widetilde{L}_Q(\boldsymbol{\beta})$ denotes the quadratic approximation of the log-likelihood. 
This quadratic approximation is defined by
\begin{equation}\label{eq:LQtilde}
-\widetilde{L}_Q(\boldsymbol{\beta})=\frac12\|\mathcal{Y}-\mathcal{X}\boldsymbol{\beta}\|_2^2,
\end{equation}
where
\begin{equation}\label{eq:def_Y_X}
\mathcal{Y}=\Lambda^{1/2}U'\boldsymbol{\beta}^{(0)}
+\Lambda^{-1/2}U'\left(\frac{\partial L}{\partial \boldsymbol{\beta}}(\boldsymbol{\beta}^{(0)},\widehat{\boldsymbol{\gamma}}, \alpha^{(0)})\right)' ,\;  \mathcal{X}=\Lambda^{1/2}U',
\end{equation}
with $\|\cdot\|_2$ denoting the $\ell_2$ norm in $\mathbb{R}^{p+1}$. The detailed computations for obtaining the expression \eqref{eq:LQtilde} of $\widetilde{L}_Q(\boldsymbol{\beta})$ are provided in Section \ref{sub:var_sec}.

\subsubsection{Estimation of $\alpha^\star$}\label{sec:estim_alpha}

To estimate $\alpha^\star$ we shall use a maximum likelihood approach in the classical GLM model, as described in~\cite{piegorsch1990}, meaning that in \eqref{eq:mut_Wt} the ARMA part is ignored. In the GLM model we take the design matrix $X$ composed of regressor variables $x_{t,i}$, for $1 \leq t \leq n$ and $i$ such that the corresponding $\hat{\beta}_i$ was estimated to be non-null in the variable selection step.

\subsubsection{Stability selection}\label{subsec:stab_sel}
In order to guarantee the robustness of the selected variables, we use the stability selection approach by \cite{meinshausen:buhlmann:2010} for obtaining the final estimator $\widehat{\pmb{\beta}}$ of $\pmb{\beta^{\star}}$. The idea of stability selection is the following.
The vector $\mathcal{Y}$ defined in (\ref{eq:def_Y_X}) is randomly split into a number of subsamples of size $(p+1)/2$, corresponding to half of the length of $\mathcal{Y}$. In our numerical experiments the number of subsamples is equal to 1000. For each subsample $\mathcal{Y}^{(s)}$ and the corresponding design matrix $\mathcal{X}^{(s)}$, we apply Criterion \eqref{eq:beta_hat} with a given $\lambda$ and by replacing $\mathcal{Y}$ and $\mathcal{X}$ with $\mathcal{Y}^{(s)}$ and $\mathcal{X}^{(s)}$, respectively. For each subsampling, we store the indices $i$ of the non-null $\widehat{\beta}_{i}$.
In the end, we calculate the frequency of index selection, namely the number of times each $i$ was selected divided by the number of subsamples. For a given threshold, in the final set of selected variables, we keep the ones whose indices have a frequency larger than this threshold. Concerning the choice of $\lambda$, we consider  the smallest element of the grid of $\lambda$ provided by the R \texttt{glmnet} package, called $\mathsf{ss\_min}$ in the following. It is also possible to use the $\lambda$ obtained by cross-validation (Chapter 7 of \cite{hastie2009elements}), called $\mathsf{ss\_cv}$ in the following.

\subsection{Description of the algorithm}\label{subsec:algo}
The algorithmic implementation of the methodology can be summarised as follows:
\begin{itemize}
\item\textsf{Initialization.} For $\boldsymbol{\beta}^{(0)}$ we take the estimator of $\boldsymbol{\beta}^\star$ obtained by fitting a GLM to the observations $Y_{1},\dots,Y_{n}$, thus ignoring the ARMA part of the model. For $\alpha^{(0)}$, we take the ML estimate of $\alpha^{\star}$ of the same GLM model. For $\boldsymbol{\gamma}^{(0)}$, we take the null vector.
\item\textsf{Newton-Raphson algorithm.} We use the recursion defined in \eqref{eq:newton_raphson:gamma} with
the initialization $(\boldsymbol{\beta}^{(0)},\boldsymbol{\gamma}^{(0)}, \alpha^{(0)})$ obtained in the previous step and
we stop at the iteration $R$ such that $\|\boldsymbol{\gamma}^{(R)}-\boldsymbol{\gamma}^{(R-1)}\|_\infty<10^{-6}$.
\item\textsf{Variable selection.} To obtain a sparse estimator of $\boldsymbol{\beta}^\star$, we use Criterion \eqref{eq:beta_hat},
  where $\boldsymbol{\beta}^{(0)}$, $\widehat{\boldsymbol{\gamma}}$, and $\alpha^{(0)}$ appearing in \eqref{eq:Ltilde} are replaced by $\boldsymbol{\beta}^{(0)}$, $\boldsymbol{\gamma}^{(R)}$, and $\alpha^{(0)}$
obtained in the previous steps. We get the indices $i$ by using the stability selection approach described in Section \ref{subsec:stab_sel}.
\item\textsf{Reestimation.} We fit a GLM to the observations $Y_{1},\dots,Y_{n}$ and the design matrix $X$, in which we leave only the columns corresponding to the indices $i$ that we got in the previous step. We obtain $\widehat{\boldsymbol{\beta}}$ and $\widehat{\alpha}$ as the final estimates of $\boldsymbol{\beta}^\star$ and $\alpha^{\star}$.
\end{itemize}

This procedure can be improved by iterating the \textsf{Newton-Raphson algorithm}, \textsf{Variable selection}, and \textsf{Reestimation} steps. More precisely, let us denote by $\widehat{\boldsymbol{\beta}}_1$, $\boldsymbol{\gamma}^{(R_1)}$, and $\widehat{\alpha}_1$ the values of $\widehat{\boldsymbol{\beta}}$, $\boldsymbol{\gamma}^{(R)}$,  $\widehat{\alpha}$ obtained in the four steps described above at the first iteration. At the second iteration, we replace $(\boldsymbol{\beta}^{(0)},\boldsymbol{\gamma}^{(0)}, \alpha^{(0)})$  appearing in the \textsf{Newton-Raphson algorithm} step with $(\widehat{\boldsymbol{\beta}}_1,\boldsymbol{\gamma}^{(R_1)}, \widehat{\alpha}_1)$ and continue the steps. At the end of this second iteration, $\widehat{\boldsymbol{\beta}}_2$, $\boldsymbol{\gamma}^{(R_2)}$ and $\widehat{\alpha}_2$ denote the obtained values of $\widehat{\boldsymbol{\beta}}$, $\boldsymbol{\gamma}^{(R)}$, and $\widehat{\alpha}$, respectively. This approach is iterated until the stabilisation of $\boldsymbol{\gamma}^{(R_k)}$.

\section{Numerical experiments}\label{sec:num}

In this section we study the performance of our method, which is implemented in the R package \texttt{NBtsVarSel} available on the CRAN (Comprehensive R Archive Network), using synthetic data generated from the model defined by \eqref{eq:Yt}, \eqref{eq:mut_Wt} and \eqref{eq:Zt}. We study its performance in terms of support recovery, which is the identification of the non null coefficients of $\pmb{\beta}^\star$, and the estimation of $\pmb{\gamma}^\star$ and $\alpha^\star$. We generate observations $Y_1,\dots,Y_n$ satisfying the model
in \eqref{eq:Yt}, \eqref{eq:mut_Wt} and \eqref{eq:Zt} with covariates chosen in a Fourier basis defined by $x_{t,i}=\cos(2 \pi i t f/n)$, when $i=1, \ldots, [p/2]$ and $x_{t,i} = \sin(2\pi i t f/n)$, when $i=[p/2] + 1, \ldots,  p$, with $t = 1, \ldots, n$ and $f=0.7$, where $[x]$ denotes the integer part of $x$.

 We consider different settings, where we vary the number of observations $n$ and $q$, namely the length of the $\pmb{\gamma}^\star$ vector.  More precisely, in our experiments $n$ takes values in $\{150, 250, 500, 1000\}$ and $q$ in $\{1,2\}$. When $q=1$, $\gamma^\star=0.5$ and when $q=2$, $\pmb{\gamma}^\star=(0.5, 0.25)$. The value of $p$ is fixed to be $100$ with $5\%$ sparsity level (only $5\%$ of the coefficients in $\pmb{\beta}^\star$ is not zero). The non-null values of $\pmb{\beta^{\star}}$ range from $-0.64$ to $1.73$. We take $\alpha^\star = 2$, in order to ensure that the standard deviation of the observations is much larger than the mean. In each setting we performed 10 simulations with 4 iterations of the algorithm. In the following, we shall see that the estimation results stabilise starting from the second iteration. Hence there is no need to have more than four iterations.

\subsection{Estimation of the support of $\boldsymbol{\beta}^\star$}

\begin{figure}[!htbp]
  \centering
  \includegraphics[scale=0.25]{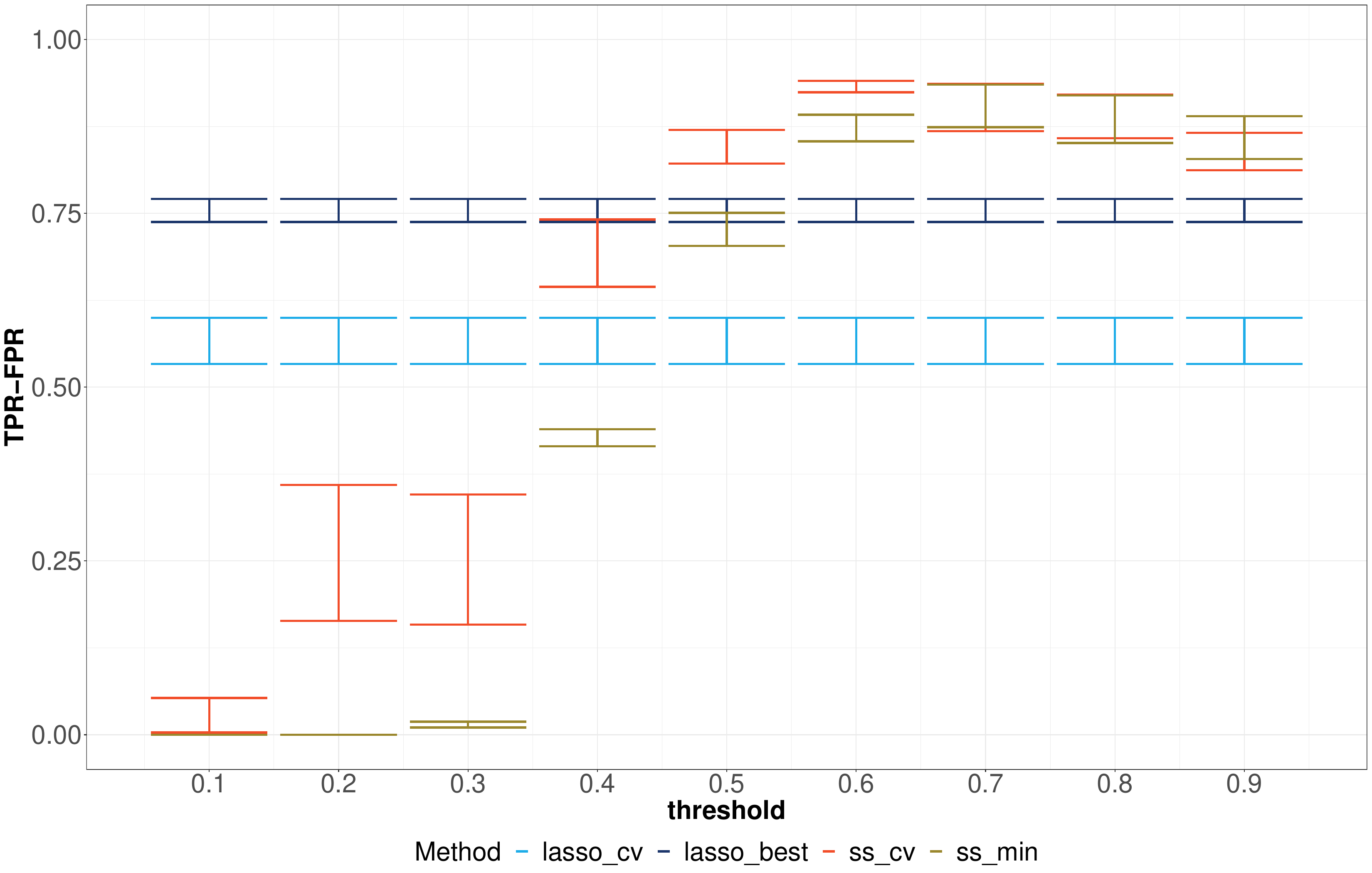}
  \caption{Error bars of the difference between the TPR and FPR associated to the support recovery of $\boldsymbol{\beta}^\star$ for four methods with respect to the thresholds when $n=1000$, $q=2$, $p=100$, $\alpha^\star=2$, and a 5\% sparsity level.
    \label{TPR_FPR_diff}}
\end{figure}

In this section, we evaluate the performance of the proposed approach in terms of support recovery of $\boldsymbol{\beta}^\star$. To do so, we calculate the  TPR (True Positive Rates, namely the proportion of non-null coefficients correctly estimated as non null) and FPR (False Positive Rates, namely the proportion of null coefficients estimated as non null). Figure \ref{TPR_FPR_diff} shows the error bars of the difference of TPR and FPR with respect to different thresholds of the stability selection step presented in Section \ref{subsec:stab_sel}. Here, we consider both the estimation with $\mathsf{ss\_min}$ and $\mathsf{ss\_cv}$. Additionally, we perform variable selection with the classical Lasso approach proposed by \cite{friedman:hastie:tibshirani:2010}. As for the $\lambda$ parameter of Lasso, we either take the $\lambda$ of standard cross-validation ($\mathsf{lasso\_cv}$) or the $\lambda$ that maximises the difference between TPR and FPR ($\mathsf{lasso\_best}$). Note that in practice it is impossible to obtain the results of $\mathsf{lasso\_best}$.

From Figure \ref{TPR_FPR_diff} we can see that our approach, both with $\mathsf{ss\_min}$ and $\mathsf{ss\_cv}$, outperforms $\mathsf{lasso\_cv}$ and $\mathsf{lasso\_best}$ when the threshold is $0.6$ and larger. In particular, the best result of $\mathsf{ss\_min}$ and $\mathsf{ss\_cv}$ are reached with the threshold $0.7$ and $0.6$, respectively. This figure presents results only in one simulation setting that we considered. The averages of the differences of TPR and FPR with corresponding standard deviations in all other settings are presented in Table \ref{TPF_FPR_difference_table}. Here, for each dataset we show the results obtained with the threshold for which the difference of TPR and FPR is the largest. Similar to Figure \ref{TPR_FPR_diff}, in all datasets $\mathsf{ss\_min}$ and $\mathsf{ss\_cv}$ give better results than $\mathsf{lasso\_cv}$ and $\mathsf{lasso\_best}$. Although the results of $\mathsf{ss\_min}$ and $\mathsf{ss\_cv}$ are quite similar, in the majority of cases $\mathsf{ss\_cv}$ gives slightly better results than $\mathsf{ss\_min}$. Hence, in the study of estimation of other parameters we will focus on the results of $\mathsf{ss\_cv}$.

\begin{table}[]
\begin{tabular}{|c|c|cc|cc|c|c|}
\hline
                      &                                             & \multicolumn{2}{c|}{$\mathsf{ss\_cv}$}                                                                   & \multicolumn{2}{c|}{$\mathsf{ss\_min}$}                                                                  & $\mathsf{lasso\_cv}$                                  & $\mathsf{lasso\_best}$                                \\ \cline{3-8} 
\multirow{-2}{*}{$n$} & \multirow{-2}{*}{$q$}                       & \multicolumn{1}{c|}{TPR-FPR}                                                                       & $t$ & \multicolumn{1}{c|}{TPR-FPR}                                                                       & $t$ & TPR-FPR                                               & TPR-FPR                                               \\ \hline
150                   & \cellcolor[HTML]{FFFFFF}                    & \multicolumn{1}{c|}{\cellcolor[HTML]{FFCCC9}\begin{tabular}[c]{@{}c@{}}0.8\\ (0.12)\end{tabular}}  & 0.5 & \multicolumn{1}{c|}{\begin{tabular}[c]{@{}c@{}}0.75\\ (0.11)\end{tabular}}                         & 0.6 & \begin{tabular}[c]{@{}c@{}}0.56\\ (0.14)\end{tabular} & \begin{tabular}[c]{@{}c@{}}0.66\\ (0.1)\end{tabular}  \\ \cline{1-1} \cline{3-8} 
250                   & \cellcolor[HTML]{FFFFFF}                    & \multicolumn{1}{c|}{\cellcolor[HTML]{FFFFFF}\begin{tabular}[c]{@{}c@{}}0.77\\ (0.13)\end{tabular}} & 0.6 & \multicolumn{1}{c|}{\cellcolor[HTML]{FFCCC9}\begin{tabular}[c]{@{}c@{}}0.79\\ (0.09)\end{tabular}} & 0.7 & \begin{tabular}[c]{@{}c@{}}0.54\\ (0.16)\end{tabular} & \begin{tabular}[c]{@{}c@{}}0.64\\ (0.1)\end{tabular}  \\ \cline{1-1} \cline{3-8} 
500                   & \cellcolor[HTML]{FFFFFF}                    & \multicolumn{1}{c|}{\cellcolor[HTML]{FFCCC9}\begin{tabular}[c]{@{}c@{}}0.9\\ (0.09)\end{tabular}}  & 0.7 & \multicolumn{1}{c|}{\cellcolor[HTML]{FFFFFF}\begin{tabular}[c]{@{}c@{}}0.87\\ (0.12)\end{tabular}} & 0.7 & \begin{tabular}[c]{@{}c@{}}0.6\\ (0.11)\end{tabular}  & \begin{tabular}[c]{@{}c@{}}0.75\\ (0.08)\end{tabular} \\ \cline{1-1} \cline{3-8} 
1000                  & \multirow{-4}{*}{\cellcolor[HTML]{FFFFFF}1} & \multicolumn{1}{c|}{\cellcolor[HTML]{FFFFFF}\begin{tabular}[c]{@{}c@{}}0.92\\ (0.1)\end{tabular}}  & 0.7 & \multicolumn{1}{c|}{\cellcolor[HTML]{FFCCC9}\begin{tabular}[c]{@{}c@{}}0.94\\ (0.07)\end{tabular}} & 0.7 & \begin{tabular}[c]{@{}c@{}}0.57\\ (0.09)\end{tabular} & \begin{tabular}[c]{@{}c@{}}0.77\\ (0.04)\end{tabular} \\ \hline
150                   &                                             & \multicolumn{1}{c|}{\cellcolor[HTML]{FFFFFF}\begin{tabular}[c]{@{}c@{}}0.71\\ (0.13)\end{tabular}} & 0.5 & \multicolumn{1}{c|}{\cellcolor[HTML]{FFCCC9}\begin{tabular}[c]{@{}c@{}}0.76\\ (0.1)\end{tabular}}  & 0.6 & \begin{tabular}[c]{@{}c@{}}0.47\\ (0.1)\end{tabular}  & \begin{tabular}[c]{@{}c@{}}0.62\\ (0.08)\end{tabular} \\ \cline{1-1} \cline{3-8} 
250                   &                                             & \multicolumn{1}{c|}{\cellcolor[HTML]{FFCCC9}\begin{tabular}[c]{@{}c@{}}0.83\\ (0.12)\end{tabular}} & 0.6 & \multicolumn{1}{c|}{\cellcolor[HTML]{FFFFFF}\begin{tabular}[c]{@{}c@{}}0.82\\ (0.14)\end{tabular}} & 0.8 & \begin{tabular}[c]{@{}c@{}}0.59\\ (0.11)\end{tabular} & \begin{tabular}[c]{@{}c@{}}0.69\\ (0.1)\end{tabular}  \\ \cline{1-1} \cline{3-8} 
500                   &                                             & \multicolumn{1}{c|}{\cellcolor[HTML]{FFCCC9}\begin{tabular}[c]{@{}c@{}}0.91\\ (0.1)\end{tabular}}  & 0.7 & \multicolumn{1}{c|}{\begin{tabular}[c]{@{}c@{}}0.9\\ (0.09)\end{tabular}}                          & 0.7 & \begin{tabular}[c]{@{}c@{}}0.59\\ (0.09)\end{tabular} & \begin{tabular}[c]{@{}c@{}}0.72\\ (0.09)\end{tabular} \\ \cline{1-1} \cline{3-8} 
1000                  & \multirow{-4}{*}{2}                         & \multicolumn{1}{c|}{\cellcolor[HTML]{FFCCC9}\begin{tabular}[c]{@{}c@{}}0.93\\ (0.03)\end{tabular}} & 0.6 & \multicolumn{1}{c|}{\cellcolor[HTML]{FFFFFF}\begin{tabular}[c]{@{}c@{}}0.9\\ (0.1)\end{tabular}}   & 0.7 & \begin{tabular}[c]{@{}c@{}}0.57\\ (0.1)\end{tabular}  & \begin{tabular}[c]{@{}c@{}}0.75\\ (0.05)\end{tabular} \\ \hline
\end{tabular}
\captionof{table}{Means of the differences of TPR and FPR with corresponding standard deviations given in parenthesis associated to the support recovery of $\pmb{\beta}^\star$ for four methods, different values of $n$, $q$, $\alpha^\star =2$, $p = 100$, and 10 simulations. The column $t$ is the threshold for which the corresponding TPR and FPR are obtained. In each setting the best results are highlighted in pink.\label{TPF_FPR_difference_table}}
\end{table}

\begin{figure}[!htbp]
  \centering
  \includegraphics[scale=0.28]{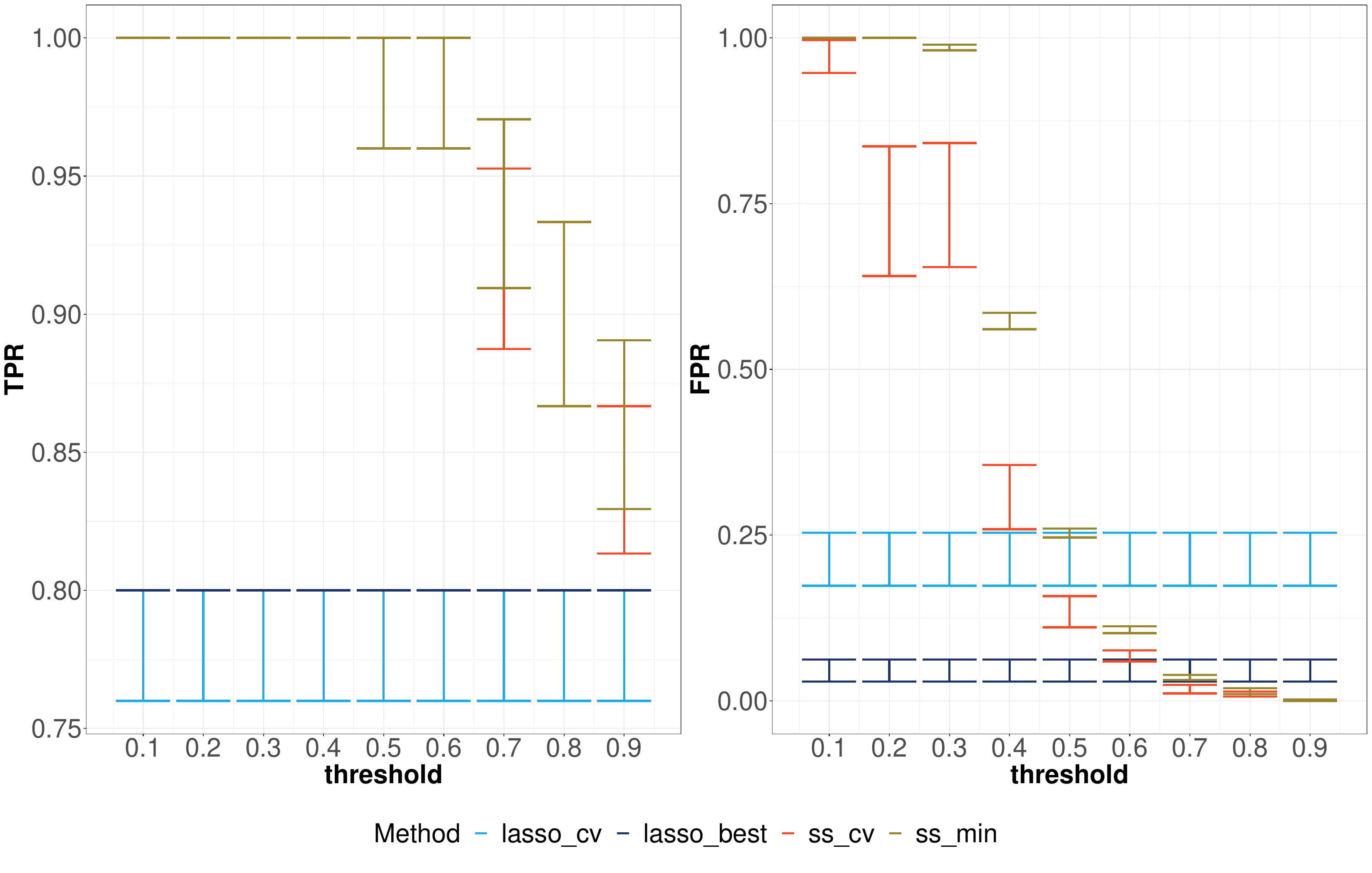}
  \caption{Error bars of the TPR and FPR associated to the support recovery of $\boldsymbol{\beta}^\star$ for four methods with respect to the thresholds when $n=1000$, $q=2$, $p=100$, $\alpha^\star=2$, and a 5\% sparsity level. 
    \label{TPR_FPR}}
\end{figure}

Depending on the application, it might be of an interest to have TPR as large as possible, or on the contrary, to minimise the FPR. Based on the objective, one can choose the optimal threshold by looking at TPR and FPR separately. In Figure \ref{TPR_FPR} we illustrate the error bars of TPR and FPR of the same dataset as in Figure \ref{TPR_FPR_diff}. For example, if the aim is to have an estimation with the smallest possible FPR, instead of taking the threshold $0.6$ in $\mathsf{ss\_cv}$ one can take the threshold $0.7$. The TPR with this threshold is still larger than the ones of $\mathsf{lasso\_cv}$ and $\mathsf{lasso\_best}$, whereas the FPR is smaller. The averages of TPR and FPR with corresponding standard deviations in all other settings are presented in Table \ref{TPF_FPR_table} in Appendix \ref{appendix_table}.

In Figure \ref{TPR_FPR_by_n} we illustrate how the error bars of the difference between the TPR and FPR depend on $n$ and $q$. As it can be expected, the methodology has better performance when there are more observations in the dataset and it has always better results than $\mathsf{lasso\_cv}$ .

\begin{figure}[!htb]
  \centering
  \includegraphics[scale=0.25]{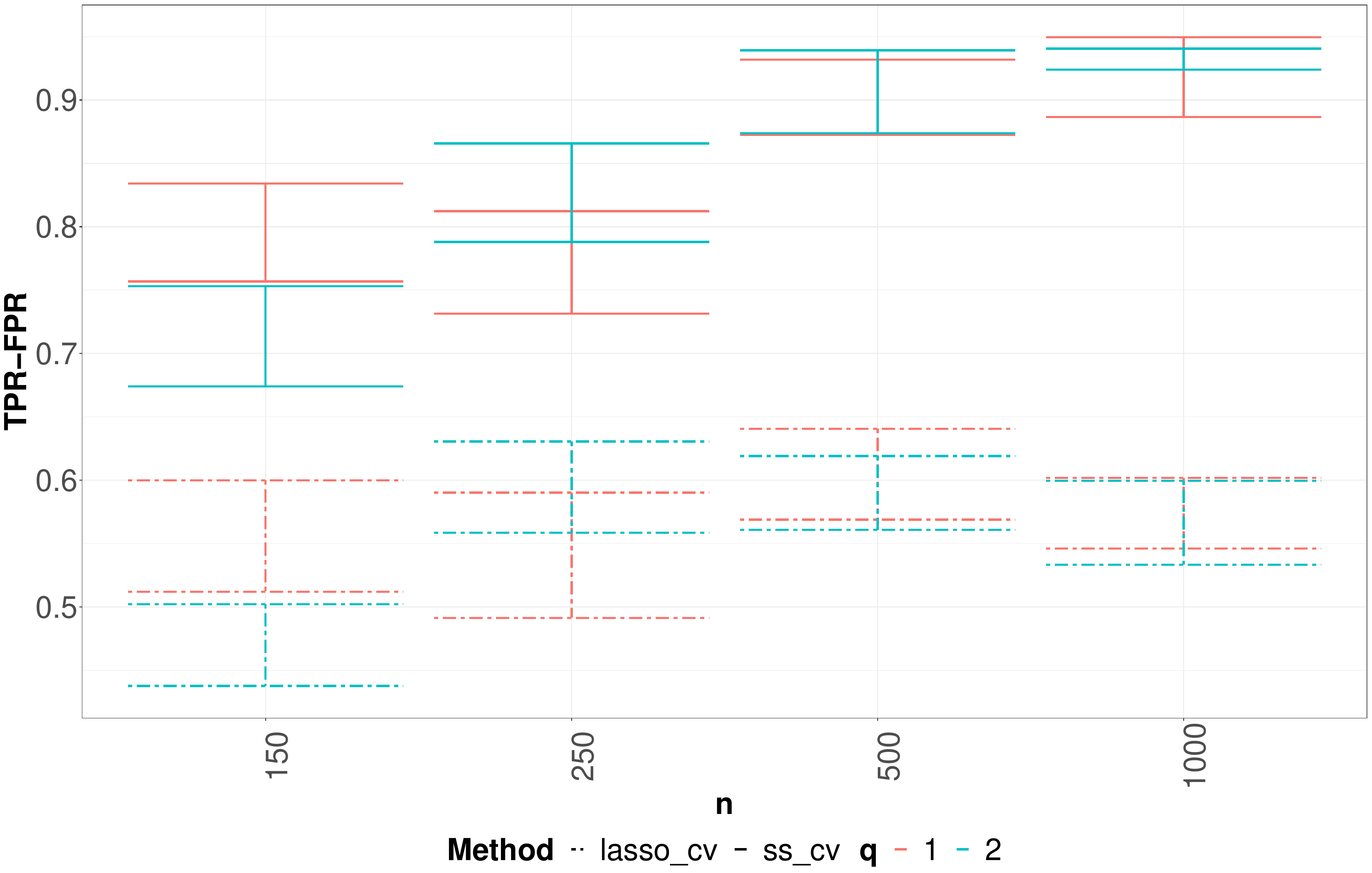}
  \caption{Error bars of the difference between the TPR and FPR associated to the support recovery of $\boldsymbol{\beta}^\star$ for $\mathsf{ss\_cv}$ and $\mathsf{lasso\_cv}$ for different values of $n$ and $q$, $p=100$, $\alpha^\star=2$, and a 5\% sparsity level.}
    \label{TPR_FPR_by_n}
\end{figure}

\clearpage

\subsection{Estimation of $\boldsymbol{\gamma}^\star$ and $\boldsymbol{\alpha}^\star$}

This section is dedicated to the estimation of $\boldsymbol{\gamma}^\star$ and $\alpha^\star$ with our methodology. All the results are obtained by the $\mathsf{ss\_cv}$ approach and in each setting we chose the threshold from Table \ref{TPF_FPR_difference_table}.

\begin{figure}[!htbp]
  \centering
  \includegraphics[scale=0.21]{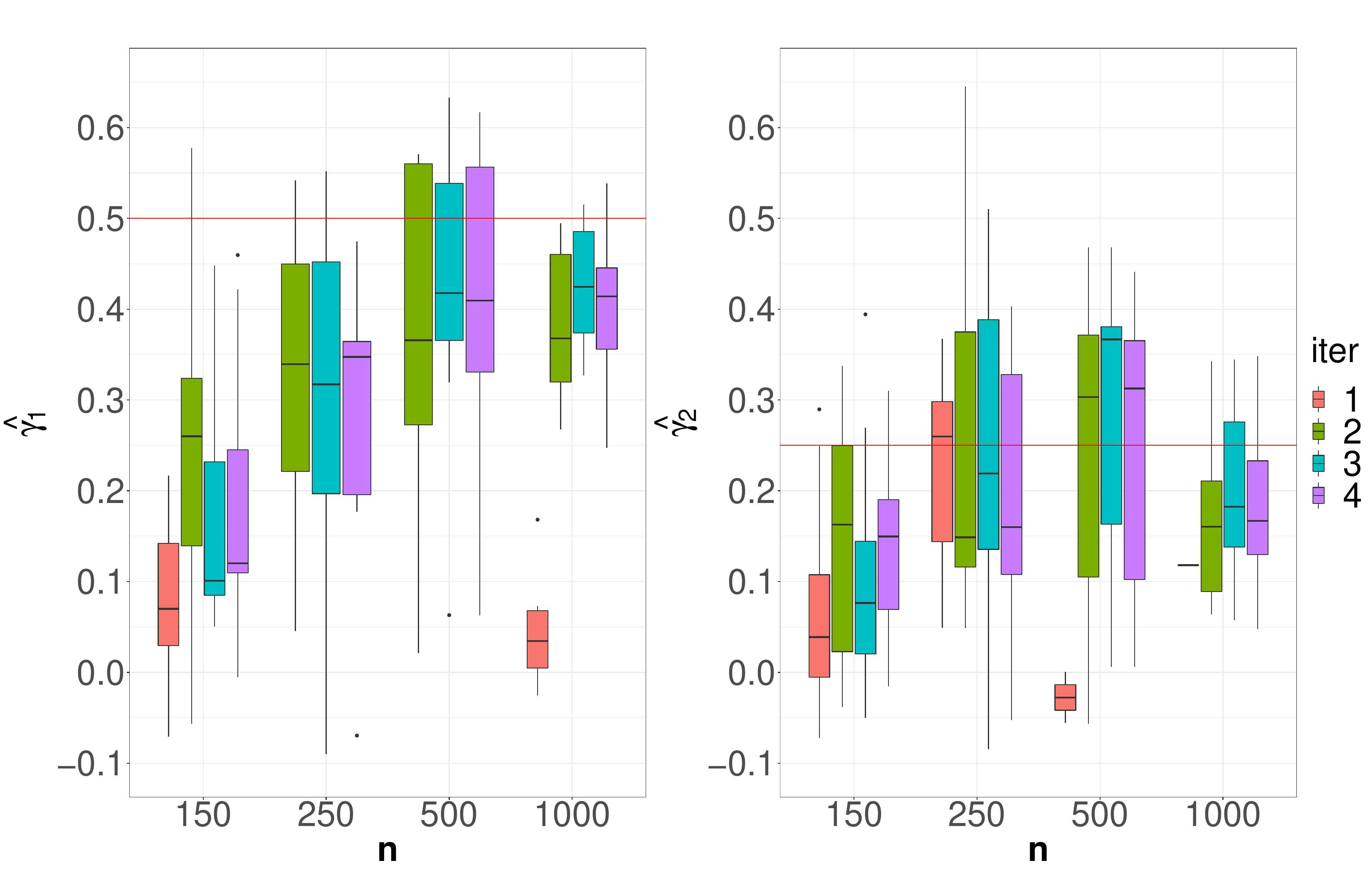}
  \caption{Boxplots for the estimations of $\boldsymbol{\gamma}^\star$ in Model \eqref{eq:mut_Wt} for $q = 2$, $p = 100$, $\alpha^\star = 2$, a 5\% sparsity level, and different values of $n$ obtained by $\mathsf{ss\_cv}$. For each $n$ the threshold is chosen corresponding to Table \ref{TPF_FPR_table}. Different colours refer to different iterations of the algorithm (\texttt{iter}). The horizontal lines correspond to the values of the $\gamma_i^\star$'s.}
    \label{gamma_est_2}
\end{figure}

Figure \ref{gamma_est_2} illustrates the impact of $n$ on the estimation of $\boldsymbol{\gamma}^\star$ when $q=2$. Similar to the results in the previous section, the estimation improves when $n$ increases and the estimations of both $\gamma_1$ and $\gamma_2$ are closer to the true values. Iterating the algorithm has positive effects: the estimation of later iterations is better than the estimation at the first iteration.

\begin{figure}[!htbp]
  \centering
  \includegraphics[scale=0.21]{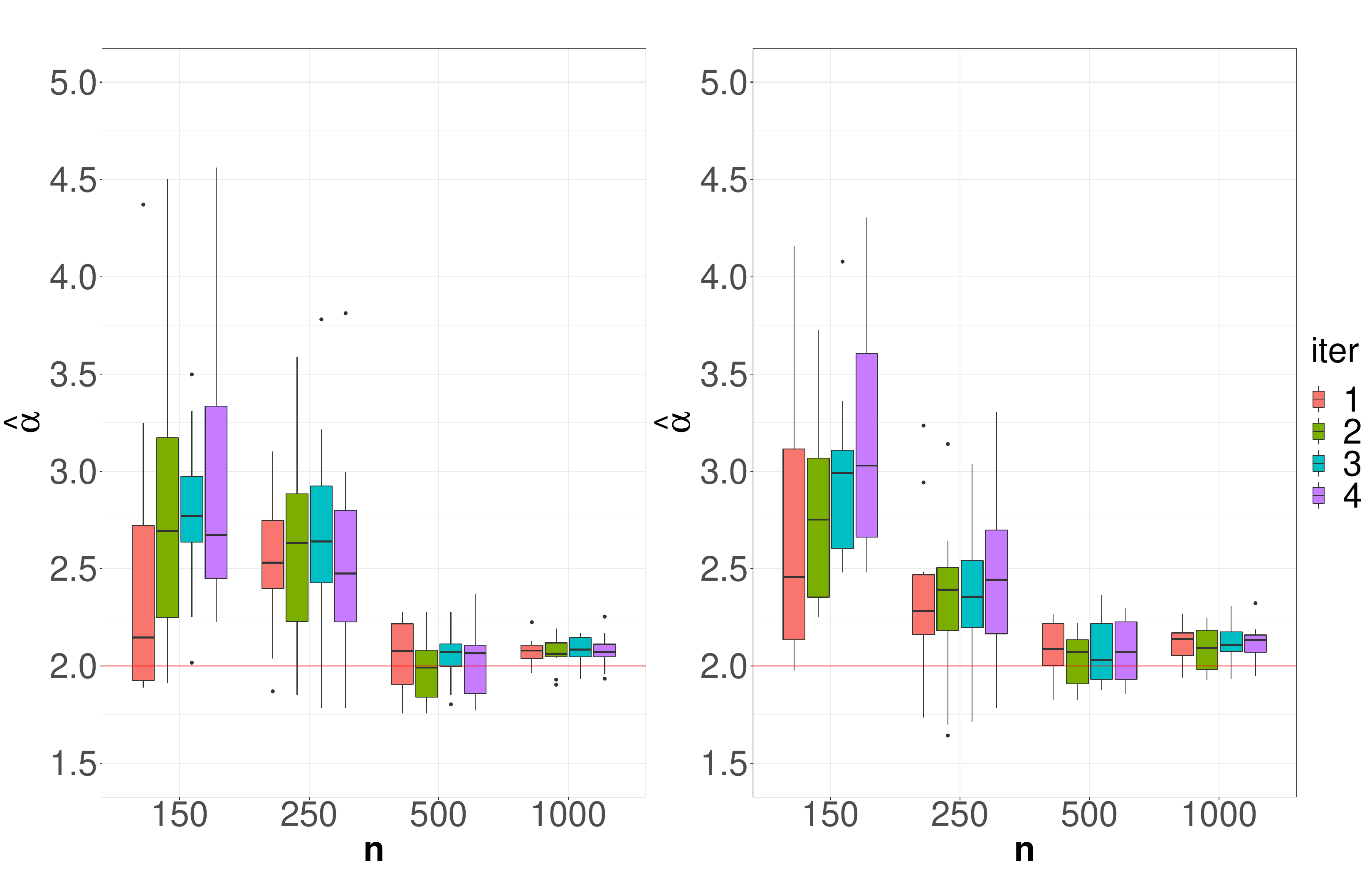}
  \caption{Boxplots for the estimations of $\alpha^\star$ in Model \eqref{eq:mut_Wt} for $p = 100$, $\alpha^\star = 2$, a 5\% sparsity level, and different values of $n$, $q=1$ (left), $q=2$ (right), and  obtained by $\mathsf{ss\_cv}$. For each $n$ the threshold is chosen corresponding to Table \ref{TPF_FPR_table}. Different colours refer to different iterations of the algorithm (\texttt{iter}). The horizontal line corresponds to the values of the $\alpha^\star$.}
    %All the $\beta_i^\star=0$ except for five of them: $\beta_1^\star=1.73$, $\beta_3^\star=0.38$, $\beta_{17}^\star=0.29$, $\beta_{33}^\star=-0.64$ and $\beta_{44}^\star=-0.13$.
    \label{alpha_est}
\end{figure}

Figure \ref{alpha_est} demonstrates the estimation of $\alpha^\star$ in the settings with two different values of $q$. While for smaller values of $n$ $\alpha^\star$ is overestimated, the results are very close to the true value for $n=500$ and $n=1000$, both for $q=1$ and $q=2$. Once again, iterating the algorithm improves and stabilises the estimation.

\clearpage

\subsection{Numerical performance}
\begin{figure}[!htbp]
\centering
\includegraphics[scale=0.25]{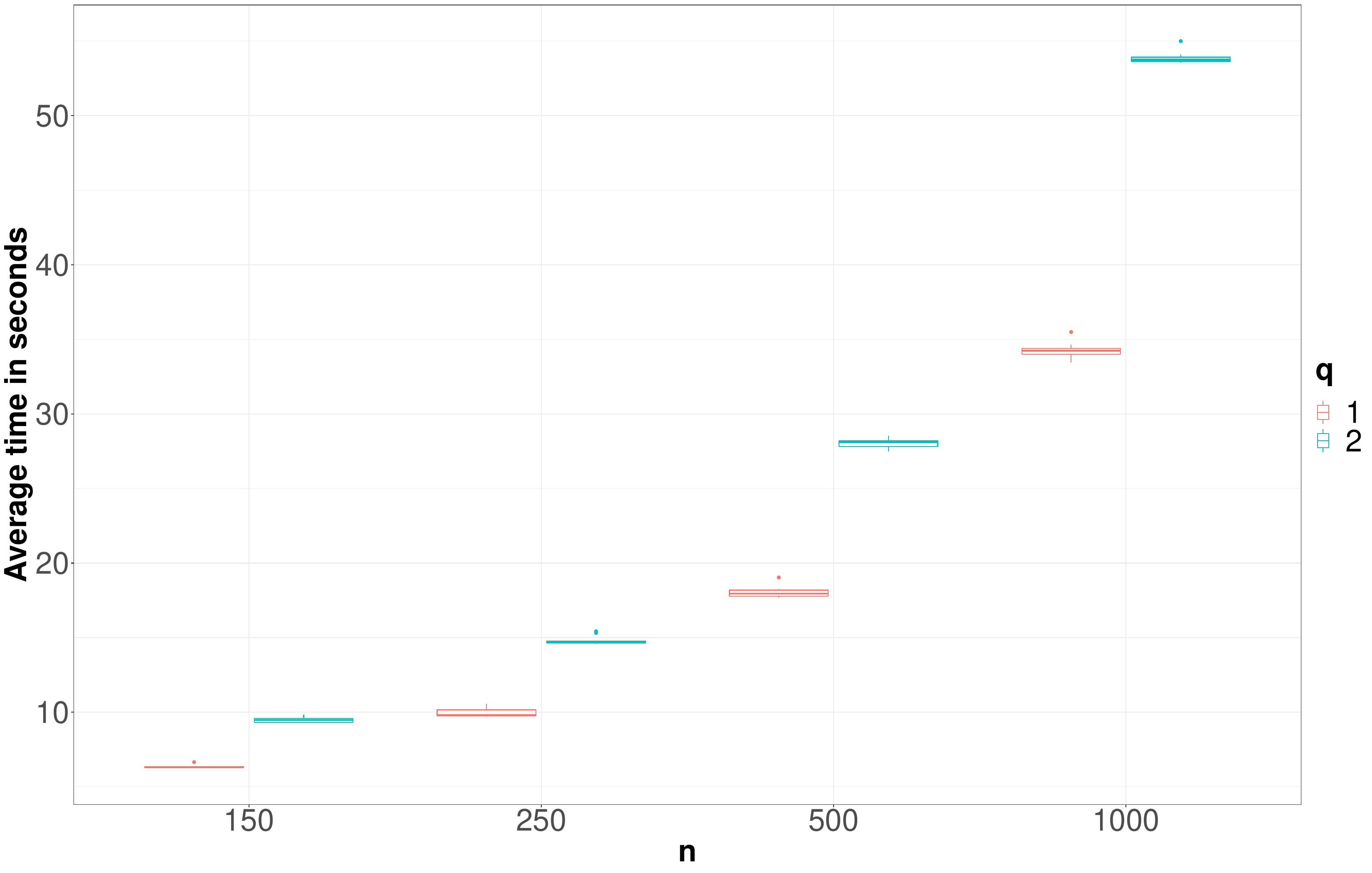}
\caption{Boxplots of the computational times in seconds  in the case where $p=100$, $\alpha^\star=2$, a 5\% sparsity level, different values of $n$ and $q$, a given threshold and one iteration.}
\label{fig:time}
\end{figure}

Figure \ref{fig:time} displays the means of the computational times of our methodology in the simulation frameworks discussed previously. We present only the results of $\mathsf{ss\_cv}$ since they are identical to the ones of $\mathsf{ss\_min}$. The timings were obtained on a workstation with 32GB of RAM and Intel Core i7-9700 (3.00GHz) CPU. For a given threshold and one iteration the algorithm needs less than one minute to process a dataset when  $n=1000$, $p=100$ and $q=2$. Moreover, it is slightly faster when $q$ is smaller. Clearly, when $n$ is smaller, the algorithms needs less time to execute. 

\section{Application to RNA-Seq time series data}\label{sec:application}
With RNA sequencing (RNA-Seq) it is possible to identify and count the numbers of RNA fragments present in a biological sample. Linking these RNA fragments to genes allows determining the expression level of genes as integer counts. Over the past decades, advances in RNA-Seq analysis have revealed that many eukaryotic genomes were transcribed outside of protein-coding genes. These new transcripts have been named non-coding RNAs (ncRNAs, \cite{ariel:2015}) as opposed to coding RNAs, which code for proteins. Among these ncRNAs, long non-coding RNAs (lncRNAs) are a heterogeneous group of RNA molecules regulating genome expression. The purpose of this application is to identify the lncRNAs, the expression of which affects the expression of coding genes, by using the temporal evolution of the expression of both coding genes and lncRNAs.

\begin{figure}[!htbp]
  \centering
  \includegraphics[scale=0.25]{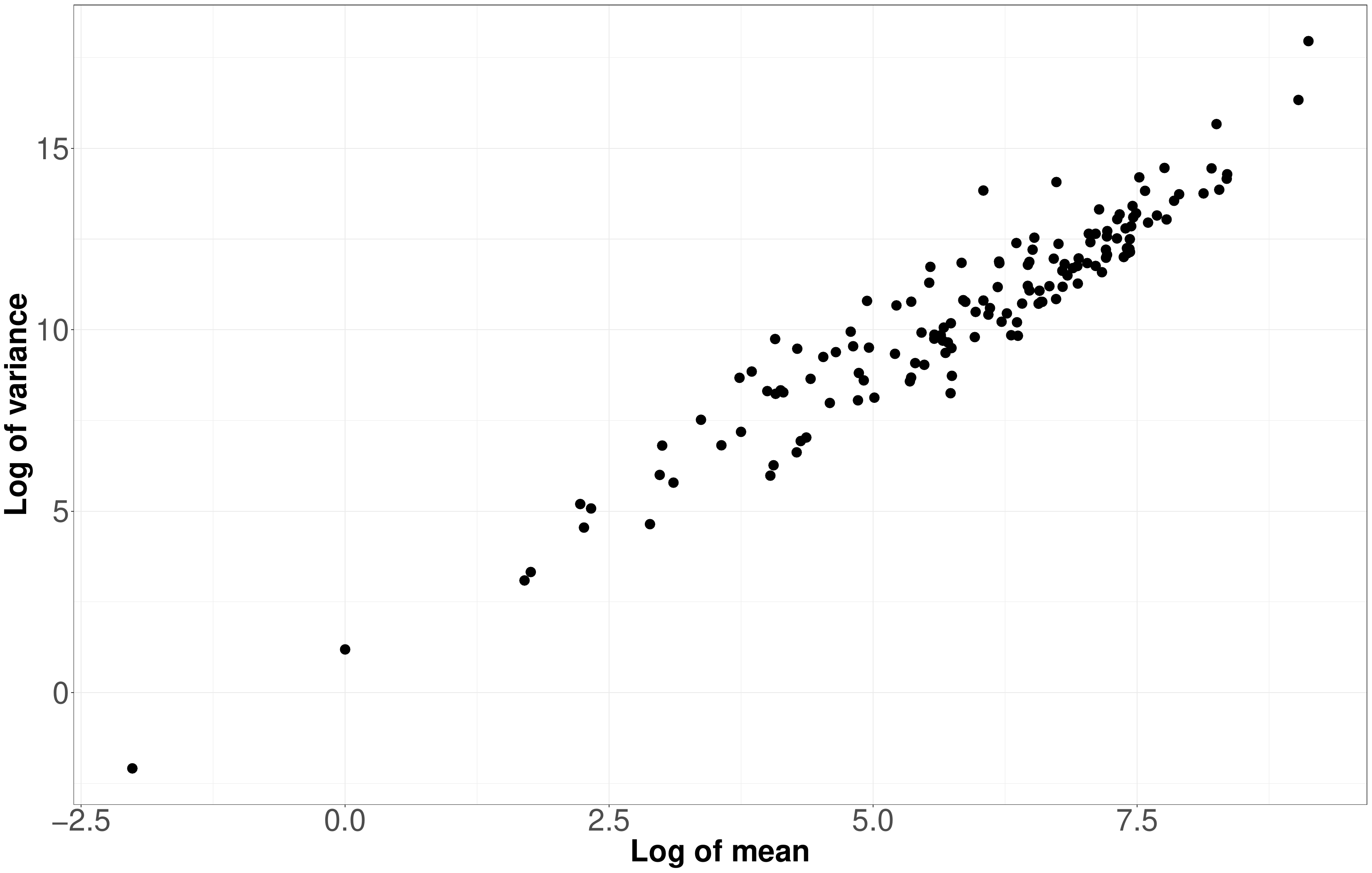}
  \caption{Scatter plot of the means and the variances of 145 RNA-Seq time series.
    \label{mean_var_real_data}}
\end{figure}

For the application of our methodology, we consider 145 RNA-Seq time series of coding genes each having a length $n=15$. The purpose of the application is to find which lncRNAs among $p=95$ affect the expression values of coding genes. Figure \ref{mean_var_real_data} shows the relation between the log of the mean and the log of the variance of each RNA-Seq time series. As it can be seen, the variances of the observations are much larger than their means. In addition, the expression of coding genes are integer-valued, therefore we are modelling the RNA-Seq time series with a negative binomial GLARMA model. Strictly speaking, for each coding gene, the time series is described by its expression (values) at 15 temporal points. In Model \eqref{eq:Yt}, \eqref{eq:mut_Wt}, and \eqref{eq:Zt} the expression of a given coding gene at time $t$ is denoted by $Y_t$ with $t=1, 2, \ldots, n =15$ and the expression of the $j$th lncRNAs at time $t$ is denoted by $x_{j,t}$ with $j=1, 2, \ldots, p =95$. Our goal is to find which lncRNAs affect the values of $(Y_t)$ for each coding gene. In other words, we aim at finding which $\beta_k^\star$ are non null.

\subsection{Choice of the threshold}
In this section we conduct additional experiments for choosing the threshold of $\mathsf{ss \_ cv}$ in our methodology. We consider simulated data in the specific context of this application with $n=15$ and $p=95$. We take the $x_{j,t}$ corresponding to the gene expression data of the lncRNAs and generate the $Y_t$'s by the model described in \eqref{eq:Yt}, \eqref{eq:mut_Wt}, and \eqref{eq:Zt} with $q=1$, $\gamma_1^\star=0.5$, $\alpha^\star = 2$ and 5 non null coefficients in $\pmb{\beta}^\star$. 

\begin{figure}[!htbp]
  \centering
  \includegraphics[scale=0.25]{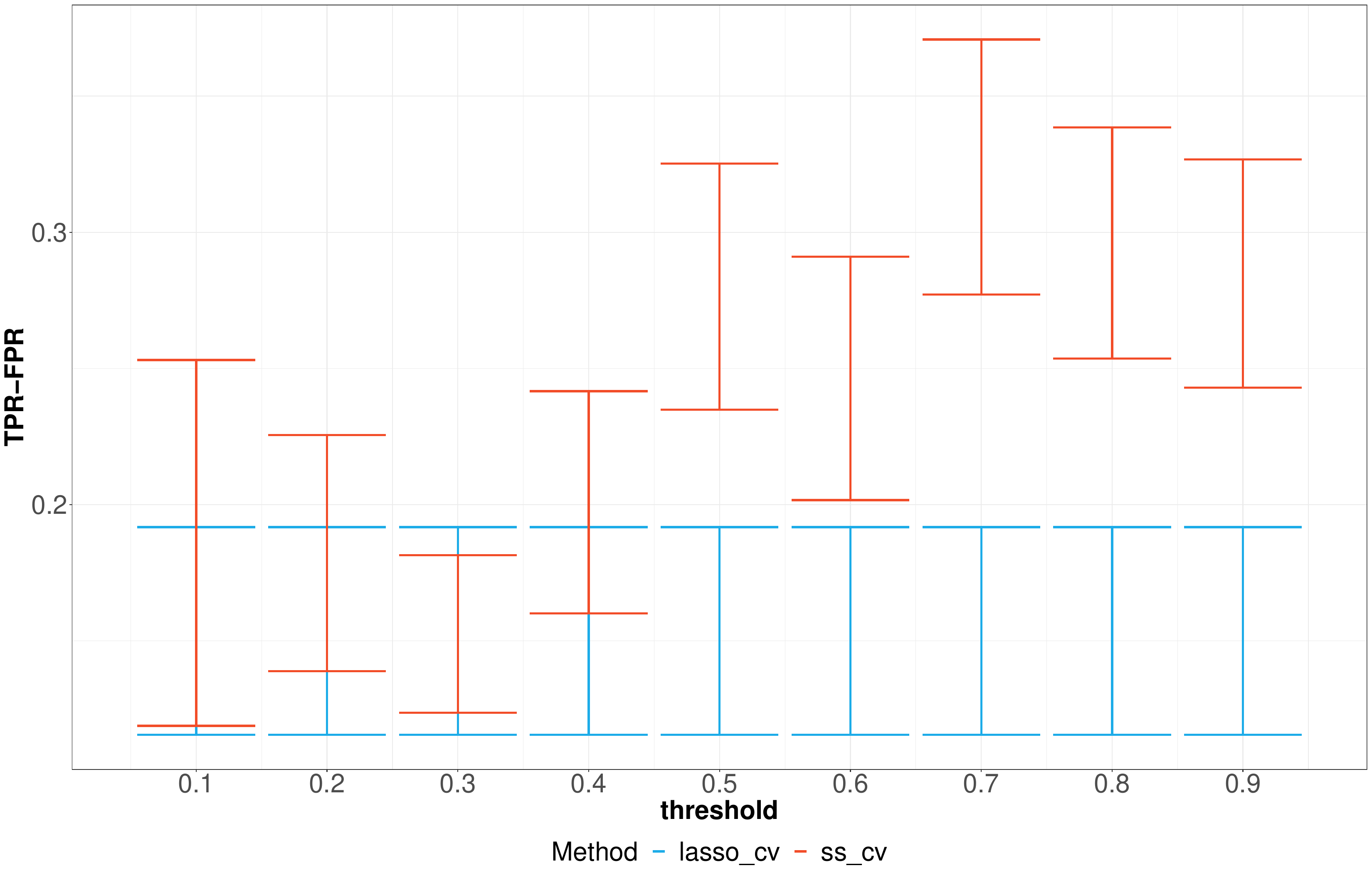}
  \caption{Difference between error bars of the TPR and FPR associated to the support recovery of $\boldsymbol{\beta}^\star$ obtained by $\mathsf{ss\_cv}$ and $\mathsf{lasso\_cv}$ with respect to the thresholds when $n=15$, $q=1$, $p=95$, $\alpha^\star=2$, and a 5 non-null coefficients.
    \label{apli_diff_TPR_FPR}}
\end{figure}

From Figure \ref{apli_diff_TPR_FPR} we can see that for the thresholds $0.5$ and larger $\mathsf{ss\_cv}$ outperforms $\mathsf{lasso\_cv}$
even in this high-dimensional framework with $n$ being much smaller than $p$. The best results are obtained with the threshold 0.7. Hence, in the application we shall use this value.

%\begin{figure}[!htbp]
 % \centering
 % \includegraphics[scale=0.28]{../code/real data/error_bars_s5.pdf}
 % \caption{Error bars of the TPR and FPR associated to the support recovery of $\boldsymbol{\beta}^\star$ obtained by $\mathsf{ss\_cv}$ and $\mathsf{lasso\_cv}$ with respect to the thresholds when $n=15$, $q=1$, $p=95$, $\alpha^\star=2$, and a 5 non-null coefficients.
%    \label{apli_TPR_FPR}}
%\end{figure}

\subsection{Results}

\begin{figure}[!htbp]
  \centering
  \includegraphics[scale=0.25]{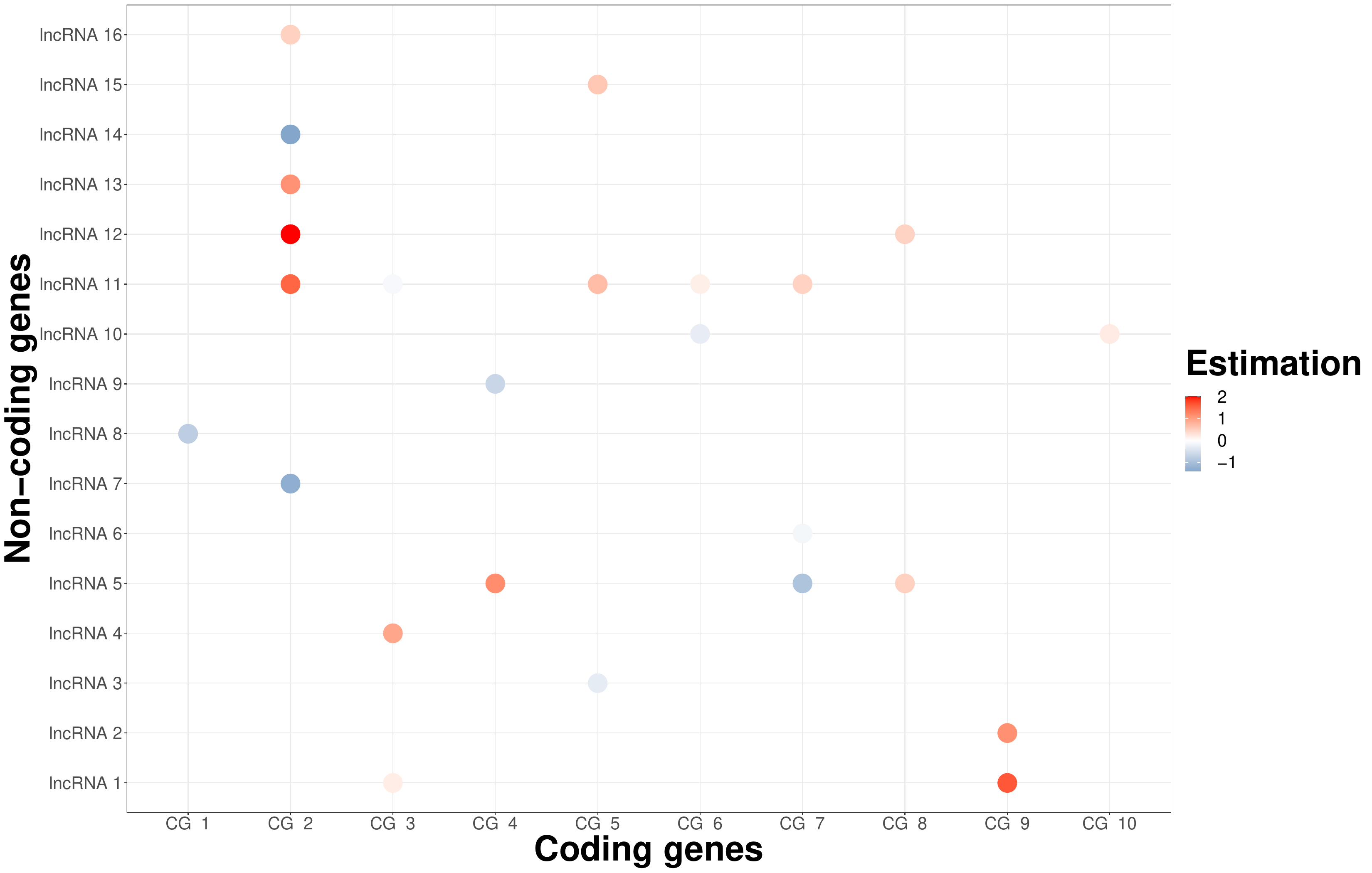}
  \caption{Estimation of $\pmb{\beta^\star}$ with $\mathsf{ss \_ cv}$ for explaining the values of 10 coding genes ($Y_t$) by some of the lncRNAs ($x_{t,i}$).
    \label{appli_var_sel}}
\end{figure}

In Figure \ref{appli_var_sel} we present results for a sample of 10 coding genes. Our method selected 16 lncRNAs out of 95 as being relevant for explaining the expression of these 10 coding genes. In this figure a dot signifies the effect of the associated lncRNA on a given coding gene. That is, the coefficient $\beta_k^\star$ corresponding to the lncRNA is estimated as non null. If the influence of a lncRNA on a given coding gene is negative, the dot is blue and if it is positive, the dot is red. The brighter the colour of the dot, the larger is the influence. For the 145 coding genes, there are in total 37 lncRNAs selected to be relevant. %If instead of considering the negative binomial model we consider the Poisson model, the method selects 93 lncRNAs. All the selected 37 lncRNAs of the negative binomial model coincide with the lncRNAs selected by the Poisson model.

\begin{figure}[!htbp]
  \centering
  \includegraphics[scale=0.25]{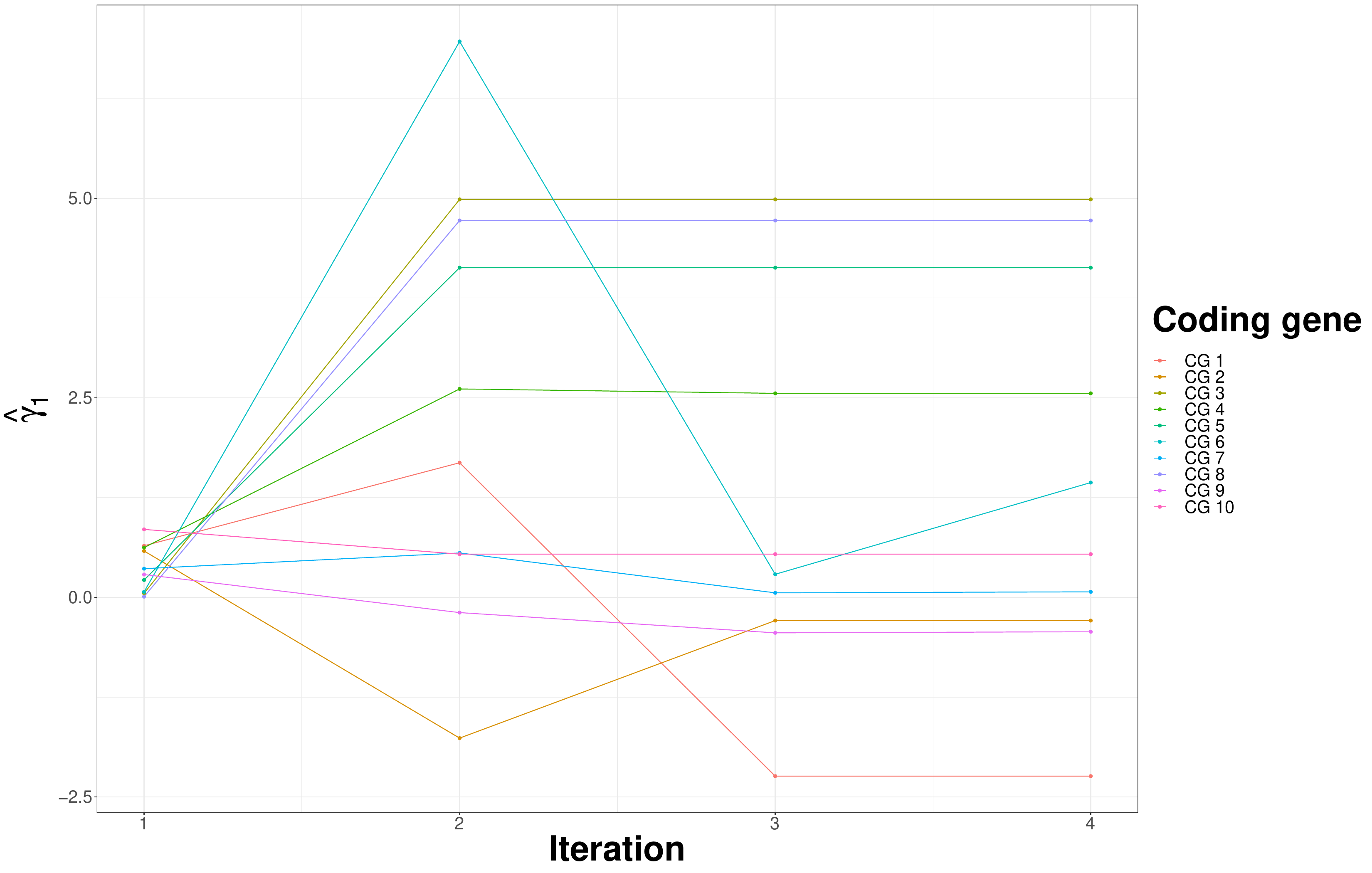}
  \caption{Estimation of $\pmb{\gamma^\star}$ with $\mathsf{ss \_ cv}$ for explaining the values of 10 coding genes ($Y_t$) by some of the lncRNAs ($x_{t,i}$).
    \label{appli_gamma_est}}
\end{figure}

Figure \ref{appli_gamma_est} displays the estimation of $\gamma_1^{\star}$ obtained for the 10 series associated to the coding genes. We take $q=1$ (number of parameters in $\pmb{\gamma}^{\star}$) since $n$ is very small and it is unrealistic to expect better results for a larger $q$. After 4 iterations for the $10$ coding genes, all the estimates of $\gamma_1^{\star}$ converge to a value in the interval from $-2.5$ to $5$.

\section{Acknowledgements}
I would like to thank my PhD supervisors Céline Lévy-Leduc, Sarah Ouadah and Laure Sansonnet for their guidance and valuable comments. I am very grateful for their help, without which this work would not have been possible.

\appendix

\section{Detailed computations}\label{sec:proofs}

\subsection{\textcolor{black}{Computation of the first and second derivatives of $W_t$ defined in (\ref{eq:Wt})}}

\subsubsection{\textcolor{black}{Computation of the first derivatives of $W_t$ }}\label{subsub:first_derive}

By the definition of $W_t$ given in (\ref{eq:Wt}), we get
\begin{equation*}
\frac{\partial W_t}{\partial \boldsymbol{\delta}}(\boldsymbol{\delta})=\frac{\partial\boldsymbol{\beta}' x_t}{\partial \boldsymbol{\delta}}+\frac{\partial Z_t}{\partial \boldsymbol{\delta}}
(\boldsymbol{\delta}),
\end{equation*}
where $\boldsymbol{\beta}$, $x_t$ and $Z_t$ are defined in (\ref{eq:Wt}). First we will calculate the derivatives of $E_t$ defined in \eqref{eq:Et}. More precisely, for all $k\in\{0,\dots,p\}$, $\ell\in\{1,\dots,q\}$ and $t\in\{1,\dots,n\}$
\begin{align*}
\frac{\partial E_t}{\partial \beta_k}&= \left( -Y_t \frac{\partial W_t}{\partial \beta_k} \exp(-W_t) \right) \cdot \frac{1}{1 + \frac{\exp(W_t)}{\alpha}}  \nonumber \\
&- \left( Y_t \exp(-W_t) - 1\right) \frac{\partial W_t}{\partial \beta_k} \cdot \exp(W_t) \cdot \frac{1}{\alpha\left(1 + \frac{\exp(W_t)}{\alpha}\right)^2} \nonumber \\
&= \left( -E_t - \frac{1}{1 + \frac{\exp(W_t)}{\alpha}} - \frac{E_t \frac{\exp(W_t)}{\alpha}}{1 + \frac{\exp(W_t)}{\alpha}} \right) \frac{\partial W_t}{\partial \beta_k} = - \left( E_t + \frac{1 + E_t \frac{\exp(W_t)}{\alpha}}{1 + \frac{\exp(W_t)}{\alpha}} \right) \frac{\partial W_t}{\partial \beta_k},  \nonumber \\
\end{align*}
\begin{align*}
\frac{\partial E_t}{\partial \gamma_\ell}&= \left( -Y_t \frac{\partial W_t}{\partial \gamma_\ell} \exp(-W_t) \right) \cdot \frac{1}{1 + \frac{\exp(W_t)}{\alpha}}  \nonumber \\
&- \left( Y_t \exp(-W_t) - 1\right) \frac{\partial W_t}{\partial \gamma_\ell} \cdot \exp(W_t) \cdot \frac{1}{\alpha\left(1 + \frac{\exp(W_t)}{\alpha}\right)^2} \nonumber \\
&= \left( -E_t - \frac{1}{1 + \frac{\exp(W_t)}{\alpha}} - \frac{E_t \frac{\exp(W_t)}{\alpha}}{1 + \frac{\exp(W_t)}{\alpha}} \right) \frac{\partial W_t}{\partial \gamma_\ell} = - \left( E_t + \frac{1 + E_t \frac{\exp(W_t)}{\alpha}}{1 + \frac{\exp(W_t)}{\alpha}} \right) \frac{\partial W_t}{\partial \gamma_\ell},  \nonumber \\
\end{align*}
and thus
\begin{align}\label{eq:gradW_beta}
\frac{\partial W_t}{\partial \beta_k}&=x_{t,k}+\frac{\partial Z_t}{\partial \beta_k}=x_{t,k}+\sum_{j=1}^{q\wedge (t-1)}\gamma_j\frac{\partial E_{t-j}}{\partial \beta_k}\nonumber\\
&=x_{t,k}-\sum_{j=1}^{q\wedge (t-1)}\gamma_j\left(E_{t-j} + \frac{1 + E_{t-j} \frac{\exp(W_{t-j})}{\alpha}}{1 + \frac{\exp(W_{t-j})}{\alpha}}\right)\frac{\partial W_{t-j}}{\partial \beta_k},\\
\frac{\partial W_t}{\partial \gamma_\ell}&= E_{t-\ell} +\frac{\partial Z_t}{\partial \gamma_\ell} = E_{t-\ell}+\sum_{j=1}^{q\wedge (t-1)} \gamma_j\frac{\partial E_{t-j}}{\partial\gamma_\ell}\nonumber\\\label{eq:gradW_gamma}
&=E_{t-\ell}-\sum_{j=1}^{q\wedge (t-1)}\gamma_j\left(E_{t-j} + \frac{1 + E_{t-j} \frac{\exp(W_{t-j})}{\alpha}}{1 + \frac{\exp(W_{t-j})}{\alpha}}\right)\frac{\partial W_{t-j}}{\partial \gamma_\ell},
\end{align}
where we used that  $E_t=0,\; \forall t\leq 0$.

The first derivatives of $W_t$ are thus obtained from the following recursive expressions. For all $k\in\{0,\dots,p\}$ 
\begin{align*}
\frac{\partial W_1}{\partial \beta_k}&=x_{1,k},\\
\frac{\partial W_2}{\partial \beta_k}&=x_{2,k}-\gamma_1\left(E_{1} + \frac{1 + E_1 \frac{\exp(W_1)}{\alpha}}{1 + \frac{\exp(W_1)}{\alpha}}\right)\frac{\partial W_{1}}{\partial \beta_k},
\end{align*}
where
\begin{equation}\label{eq:E1}
W_1=\boldsymbol{\beta}' x_1 \textrm{ and } E_1=\frac{Y_1 - \exp(W_1)}{\exp(W_1) + \exp(W_1)^2/ \alpha}.
\end{equation}
Moreover,
\begin{equation*}
\frac{\partial W_3}{\partial \beta_k}=x_{3,k}-\gamma_1\left(E_{2} + \frac{1 + E_2 \frac{\exp(W_2)}{\alpha}}{1 + \frac{\exp(W_2)}{\alpha}}\right)\frac{\partial W_{2}}{\partial \beta_k}-\gamma_2\left(E_{1} + \frac{1 + E_1 \frac{\exp(W_1)}{\alpha}}{1 + \frac{\exp(W_1)}{\alpha}}\right)\frac{\partial W_{1}}{\partial \beta_k},
\end{equation*}
where
\begin{equation}\label{eq:E2}
W_2=\boldsymbol{\beta}' x_2  +\gamma_1 E_{1},\; E_2=\frac{Y_2 - \exp(W_2)}{\exp(W_2) + \exp(W_2)^2/ \alpha},
\end{equation}
and so on. In the same way, for all $\ell\in\{1,\dots,q\}$
\begin{align*}
\frac{\partial W_1}{\partial \gamma_\ell}&=0,\\
\frac{\partial W_2}{\partial \gamma_\ell}&=E_{2-\ell},\\
\frac{\partial W_3}{\partial \gamma_\ell}&=E_{3-\ell}-\gamma_1\left(E_{2} + \frac{1 + E_2 \frac{\exp(W_2)}{\alpha}}{1 + \frac{\exp(W_2)}{\alpha}}\right)\frac{\partial W_{2}}{\partial \gamma_\ell}%=E_{3-\ell}-\gamma_1(1+E_{2})E_{2-\ell},
\end{align*}
and so on, where $E_t=0,\; \forall t\leq 0$, and $E_1$ and $E_2$ are defined in (\ref{eq:E1}) and (\ref{eq:E2}), respectively.

\subsubsection{\textcolor{black}{Computation of the second derivatives of $W_t$}}\label{subsub:second_derive}
By using (\ref{eq:gradW_beta}) and (\ref{eq:gradW_gamma}), we get that for all $j,k\in\{0,\dots,p\}$, $\ell,m\in\{1,\dots,q\}$ and $t\in\{1,\dots,n\}$,
\begin{align*}
\frac{\partial^2 W_{t}}{\partial \beta_j\partial \beta_k}&= - \sum_{i=1}^{q\wedge (t-1)}\gamma_i \left( E_{t-i} + \frac{1 + E_{t-i} \frac{\exp(W_{t-i})}{\alpha} }{1 + \frac{\exp(W_{t-i})}{\alpha}}\right) \frac{\partial^2 W_{t-i}}{\partial \beta_k \partial \beta_j} \\
&+ \sum_{i=1}^{q\wedge (t-1)}\gamma_i \left( E_{t-i} + 2 \frac{E_{t-i} \frac{\exp(2W_{t-i})}{\alpha} + Y_{t-i}}{\alpha \left( 1 + \frac{\exp(W_{t-i})}{\alpha}\right)^2} + \frac{1 - E_{t-i} \frac{\exp(W_{t-i})}{\alpha}}{1+\frac{\exp(W_{t-i})}{\alpha}} \right) \frac{\partial W_{t-i}}{\partial \beta_j} \frac{\partial W_{t-i}}{\partial \beta_k},
\end{align*}

\begin{align*}
\frac{\partial^2 W_t}{\partial \gamma_\ell\partial \gamma_m}&= \frac{\partial E_{t-\ell}}{\partial \gamma_m} -\left(E_{t-m} + \frac{1 + E_{t-m} \frac{\exp(W_{t-m})}{\alpha}}{1 + \frac{\exp(W_{t-m})}{\alpha}}\right)\frac{\partial W_{t-m}}{\partial \gamma_\ell} \\
&- \sum_{i=1}^{q\wedge (t-1)}\gamma_i \left( E_{t-i} + \frac{1 + E_{t-i} \frac{\exp(W_{t-i})}{\alpha} }{1 + \frac{\exp(W_{t-i})}{\alpha}}\right) \frac{\partial^2 W_{t-i}}{\partial \gamma_\ell \partial \gamma_m} \\
&+ \sum_{i=1}^{q\wedge (t-1)}\gamma_i \left( E_{t-i} + 2 \frac{E_{t-i} \frac{\exp(2W_{t-i})}{\alpha} + Y_{t-i}}{\alpha \left( 1 + \frac{\exp(W_{t-i})}{\alpha}\right)^2} + \frac{1 - E_{t-i} \frac{\exp(W_{t-i})}{\alpha}}{1+\frac{\exp(W_{t-i})}{\alpha}} \right) \frac{\partial W_{t-i}}{\partial \gamma_\ell} \frac{\partial W_{t-i}}{\partial \gamma_m} \\
& = - \left( E_{t-\ell} + \frac{1 + E_{t-\ell} \frac{\exp(W_{t-\ell})}{\alpha}}{1 + \frac{\exp(W_{t-\ell})}{\alpha}} \right) \frac{\partial W_{t-\ell}}{\partial \gamma_m} -\left(E_{t-m} + \frac{1 + E_{t-m} \frac{\exp(W_{t-m})}{\alpha}}{1 + \frac{\exp(W_{t-m})}{\alpha}}\right)\frac{\partial W_{t-m}}{\partial \gamma_\ell} \\
&- \sum_{i=1}^{q\wedge (t-1)}\gamma_i \left( E_{t-i} + \frac{1 + E_{t-i} \frac{\exp(W_{t-i})}{\alpha} }{1 + \frac{\exp(W_{t-i})}{\alpha}}\right) \frac{\partial W_{t-i}^2}{\partial \gamma_\ell \partial \gamma_m} \\
&+ \sum_{i=1}^{q\wedge (t-1)}\gamma_i \left( E_{t-i} + 2 \frac{E_{t-i} \frac{\exp(2W_{t-i})}{\alpha} + Y_{t-i}}{\alpha \left( 1 + \frac{\exp(W_{t-i})}{\alpha}\right)^2} + \frac{1 - E_{t-i} \frac{\exp(W_{t-i})}{\alpha}}{1+\frac{\exp(W_{t-i})}{\alpha}} \right) \frac{\partial W_{t-i}}{\partial \gamma_\ell} \frac{\partial W_{t-i}}{\partial \gamma_m}. \\
\end{align*}

To compute the second derivatives of $W_t$, we shall use the following recursive expressions for all $j,k\in\{0,\dots,p\}$
\begin{align*}
\frac{\partial^2 W_1}{\partial \beta_j\partial \beta_k}&=0,\\
\frac{\partial^2 W_2}{\partial \beta_j\partial \beta_k}&=\gamma_1 \left( E_1 + 2 \frac{E_1 \frac{\exp(2W_1)}{\alpha} + Y_1}{\alpha \left( 1 + \frac{\exp(W_1)}{\alpha}\right)^2} + \frac{1 - E_1 \frac{\exp(W_1)}{\alpha}}{1+\frac{\exp(W_1)}{\alpha}} \right) \frac{\partial W_{1}}{\partial \beta_j} \frac{\partial W_{1}}{\partial \beta_k},
\end{align*}
where $E_1$ is defined in (\ref{eq:E1}) and so on. Moreover, for all $\ell,m\in\{1,\dots,q\}$
\begin{align*}
\frac{\partial^2 W_1}{\partial \gamma_\ell\partial\gamma_m}&=0,\\
\frac{\partial^2 W_2}{\partial \gamma_\ell\partial\gamma_m}&=0
\end{align*}
and so on with $E_t=0$ for all $t\leq 0$ and the first derivatives of $W_t$ computed in (\ref{eq:gradW_gamma}).

\subsection{Computational details for obtaining Criterion (\ref{eq:beta_hat})}\label{sub:var_sec}

By \eqref{eq:Ltilde},
\begin{align*}
\widetilde{L}(\boldsymbol{\beta})=\widetilde{L}(\boldsymbol{\beta}^{(0)})+\frac{\partial L}{\partial \boldsymbol{\beta}}(\boldsymbol{\beta}^{(0)},\widehat{\boldsymbol{\gamma}}, \widehat{\alpha})
U(\boldsymbol{\nu}-\boldsymbol{\nu}^{(0)})-\frac12 (\boldsymbol{\nu}-\boldsymbol{\nu}^{(0)})' \Lambda (\boldsymbol{\nu}-\boldsymbol{\nu}^{(0)}),
\end{align*}
%\textcolor{blue}{OK pour le "-" dans l'approximation de Taylor (à rediscuter) ; il faut aussi le mettre dans la définition de $\tilde{L}(\boldsymbol{\beta})$ ci-dessus}
where $\boldsymbol{\nu}-\boldsymbol{\nu}^{(0)}=U'(\boldsymbol{\beta}-\boldsymbol{\beta}^{(0)})$.
Hence,
\begin{align*}
\widetilde{L}(\boldsymbol{\beta})&=\widetilde{L}(\boldsymbol{\beta}^{(0)})+\sum_{k=0}^p
\left(\frac{\partial L}{\partial \boldsymbol{\beta}}(\boldsymbol{\beta}^{(0)},\widehat{\boldsymbol{\gamma}}, \widehat{\alpha}) U\right)_k (\nu_k-\nu_{k}^{(0)})
-\frac12\sum_{k=0}^p\lambda_k (\nu_k-\nu_{k}^{(0)})^2\\
&=\widetilde{L}(\boldsymbol{\beta}^{(0)})-\frac12\sum_{k=0}^p\lambda_k\left(\nu_k-\nu_{k}^{(0)}-\frac{1}{\lambda_k}
\left(\frac{\partial L}{\partial \boldsymbol{\beta}}(\boldsymbol{\beta}^{(0)},\widehat{\boldsymbol{\gamma}}, \widehat{\alpha}) U\right)_k\right)^2
+\sum_{k=0}^p\frac{1}{2\lambda_k}\left(\frac{\partial L}{\partial \boldsymbol{\beta}}(\boldsymbol{\beta}^{(0)},\widehat{\boldsymbol{\gamma}}, \widehat{\alpha}) U\right)_k^2,
\end{align*}
where the $\lambda_k$'s are the diagonal terms of $\Lambda$.

Since the only term depending on $\boldsymbol{\beta}$ is the second one in the last expression of $\widetilde{L}(\boldsymbol{\beta})$,
we define $\widetilde{L}_Q(\boldsymbol{\beta})$ appearing in Criterion (\ref{eq:beta_hat}) as follows:
\begin{eqnarray*}
-\widetilde{L}_Q(\boldsymbol{\beta})&=&\frac12\sum_{k=0}^p\lambda_k\left(\nu_k-\nu_{k}^{(0)}-\frac{1}{\lambda_k}
\left(\frac{\partial L}{\partial \boldsymbol{\beta}}(\boldsymbol{\beta}^{(0)},\widehat{\boldsymbol{\gamma}}, \widehat{\alpha}) U\right)_k\right)^2\\
&=&\frac12 \left\|\Lambda^{1/2}\left(\boldsymbol{\nu}-\boldsymbol{\nu}^{(0)}-\Lambda^{-1} \left(\frac{\partial L}{\partial \boldsymbol{\beta}}(\boldsymbol{\beta}^{(0)},\widehat{\boldsymbol{\gamma}}, \widehat{\alpha}) U\right)'
\right)\right\|_2^2\\
&=&\frac12 \left\|\Lambda^{1/2}U'(\boldsymbol{\beta}-\boldsymbol{\beta}^{(0)})-\Lambda^{-1/2} U' \left(\frac{\partial L}{\partial \boldsymbol{\beta}}(\boldsymbol{\beta}^{(0)},\widehat{\boldsymbol{\gamma}}, \widehat{\alpha})\right)'
\right\|_2^2\\
&=&\frac12 \left\|\Lambda^{1/2}U'(\boldsymbol{\beta}^{(0)}-\boldsymbol{\beta})+\Lambda^{-1/2} U' \left(\frac{\partial L}{\partial \boldsymbol{\beta}}(\boldsymbol{\beta}^{(0)},\widehat{\boldsymbol{\gamma}}, \widehat{\alpha})\right)'\right\|_2^2\\
&=&\frac12\|\mathcal{Y}-\mathcal{X}\boldsymbol{\beta}\|_2^2,
\end{eqnarray*}
where
\begin{equation*}
\mathcal{Y}=\Lambda^{1/2}U'\boldsymbol{\beta}^{(0)}
+\Lambda^{-1/2}U'\left(\frac{\partial L}{\partial \boldsymbol{\beta}}(\boldsymbol{\beta}^{(0)},\widehat{\boldsymbol{\gamma}}, \widehat{\alpha})\right)' ,\;  \mathcal{X}=\Lambda^{1/2}U'.
\end{equation*}

\section{Additional results} \label{appendix_table}

% \usepackage{multirow}
\begin{table}[]
\begin{tabular}{|c|c|ccc|ccc|cc|cc|}
\hline
\multirow{2}{*}{$n$} & \multirow{2}{*}{$q$} & \multicolumn{3}{c|}{$\mathsf{ss\_cv}$}                                                                                                                        & \multicolumn{3}{c|}{$\mathsf{ss\_min}$}                                                                                                                       & \multicolumn{2}{c|}{$\mathsf{lasso\_cv}$}                                                                                          & \multicolumn{2}{c|}{$\mathsf{lasso\_best}$}                                                                                        \\ \cline{3-12} 
                     &                      & \multicolumn{1}{c|}{TPR}                                                   & \multicolumn{1}{c|}{FPR}                                                   & $t$ & \multicolumn{1}{c|}{TPR}                                                   & \multicolumn{1}{c|}{FPR}                                                   & $t$ & \multicolumn{1}{c|}{TPR}                                                   & FPR                                                   & \multicolumn{1}{c|}{TPR}                                                   & FPR                                                   \\ \hline
150                  & \multirow{4}{*}{1}   & \multicolumn{1}{c|}{\begin{tabular}[c]{@{}c@{}}0.86\\ (0.1)\end{tabular}}  & \multicolumn{1}{c|}{\begin{tabular}[c]{@{}c@{}}0.06\\ (0.04)\end{tabular}} & 0.5 & \multicolumn{1}{c|}{\begin{tabular}[c]{@{}c@{}}0.86\\ (0.13)\end{tabular}} & \multicolumn{1}{c|}{\begin{tabular}[c]{@{}c@{}}0.15\\ (0.02)\end{tabular}} & 0.6 & \multicolumn{1}{c|}{\begin{tabular}[c]{@{}c@{}}0.58\\ (0.15)\end{tabular}} & \begin{tabular}[c]{@{}c@{}}0.02\\ (0.02)\end{tabular} & \multicolumn{1}{c|}{\begin{tabular}[c]{@{}c@{}}0.72\\ (0.1)\end{tabular}}  & \begin{tabular}[c]{@{}c@{}}0.06\\ (0.06)\end{tabular} \\ \cline{1-1} \cline{3-12} 
250                  &                      & \multicolumn{1}{c|}{\begin{tabular}[c]{@{}c@{}}0.8\\ (0.09)\end{tabular}}  & \multicolumn{1}{c|}{\begin{tabular}[c]{@{}c@{}}0.03\\ (0.02)\end{tabular}} & 0.6 & \multicolumn{1}{c|}{\begin{tabular}[c]{@{}c@{}}0.84\\ (0.08)\end{tabular}} & \multicolumn{1}{c|}{\begin{tabular}[c]{@{}c@{}}0.05\\ (0.02)\end{tabular}} & 0.7 & \multicolumn{1}{c|}{\begin{tabular}[c]{@{}c@{}}0.62\\ (0.15)\end{tabular}} & \begin{tabular}[c]{@{}c@{}}0.08\\ (0.05)\end{tabular} & \multicolumn{1}{c|}{\begin{tabular}[c]{@{}c@{}}0.7\\ (0.11)\end{tabular}}  & \begin{tabular}[c]{@{}c@{}}0.06\\ (0.07)\end{tabular} \\ \cline{1-1} \cline{3-12} 
500                  &                      & \multicolumn{1}{c|}{\begin{tabular}[c]{@{}c@{}}0.86\\ (0.13)\end{tabular}} & \multicolumn{1}{c|}{\begin{tabular}[c]{@{}c@{}}0.02\\ (0.02)\end{tabular}} & 0.7 & \multicolumn{1}{c|}{\begin{tabular}[c]{@{}c@{}}0.94\\ (0.1)\end{tabular}}  & \multicolumn{1}{c|}{\begin{tabular}[c]{@{}c@{}}0.06\\ (0.02)\end{tabular}} & 0.7 & \multicolumn{1}{c|}{\begin{tabular}[c]{@{}c@{}}0.76\\ (0.08)\end{tabular}} & \begin{tabular}[c]{@{}c@{}}0.16\\ (0.1)\end{tabular}  & \multicolumn{1}{c|}{\begin{tabular}[c]{@{}c@{}}0.76\\ (0.08)\end{tabular}} & \begin{tabular}[c]{@{}c@{}}0.01\\ (0.02)\end{tabular} \\ \cline{1-1} \cline{3-12} 
1000                 &                      & \multicolumn{1}{c|}{\begin{tabular}[c]{@{}c@{}}0.94\\ (0.1)\end{tabular}}  & \multicolumn{1}{c|}{\begin{tabular}[c]{@{}c@{}}0.02\\ (0.02)\end{tabular}} & 0.7 & \multicolumn{1}{c|}{\begin{tabular}[c]{@{}c@{}}0.98\\ (0.06)\end{tabular}} & \multicolumn{1}{c|}{\begin{tabular}[c]{@{}c@{}}0.04\\ (0.01)\end{tabular}} & 0.7 & \multicolumn{1}{c|}{\begin{tabular}[c]{@{}c@{}}0.8\\ (0)\end{tabular}}     & \begin{tabular}[c]{@{}c@{}}0.23\\ (0.09)\end{tabular} & \multicolumn{1}{c|}{\begin{tabular}[c]{@{}c@{}}0.8\\ (0)\end{tabular}}     & \begin{tabular}[c]{@{}c@{}}0.03\\ (0.04)\end{tabular} \\ \hline
150                  & \multirow{4}{*}{2}   & \multicolumn{1}{c|}{\begin{tabular}[c]{@{}c@{}}0.84\\ (0.13)\end{tabular}} & \multicolumn{1}{c|}{\begin{tabular}[c]{@{}c@{}}0.08\\ (0.05)\end{tabular}} & 0.5 & \multicolumn{1}{c|}{\begin{tabular}[c]{@{}c@{}}0.9\\ (0.11)\end{tabular}}  & \multicolumn{1}{c|}{\begin{tabular}[c]{@{}c@{}}0.15\\ (0.03)\end{tabular}} & 0.6 & \multicolumn{1}{c|}{\begin{tabular}[c]{@{}c@{}}0.52\\ (0.14)\end{tabular}} & \begin{tabular}[c]{@{}c@{}}0.05\\ (0.07)\end{tabular} & \multicolumn{1}{c|}{\begin{tabular}[c]{@{}c@{}}0.68\\ (0.1)\end{tabular}}  & \begin{tabular}[c]{@{}c@{}}0.06\\ (0.05)\end{tabular} \\ \cline{1-1} \cline{3-12} 
250                  &                      & \multicolumn{1}{c|}{\begin{tabular}[c]{@{}c@{}}0.88\\ (0.14)\end{tabular}} & \multicolumn{1}{c|}{\begin{tabular}[c]{@{}c@{}}0.04\\ (0.02)\end{tabular}} & 0.6 & \multicolumn{1}{c|}{\begin{tabular}[c]{@{}c@{}}0.86\\ (0.13)\end{tabular}} & \multicolumn{1}{c|}{\begin{tabular}[c]{@{}c@{}}0.02\\ (0.02)\end{tabular}} & 0.8 & \multicolumn{1}{c|}{\begin{tabular}[c]{@{}c@{}}0.68\\ (0.14)\end{tabular}} & \begin{tabular}[c]{@{}c@{}}0.09\\ (0.05)\end{tabular} & \multicolumn{1}{c|}{\begin{tabular}[c]{@{}c@{}}0.78\\ (0.06)\end{tabular}} & \begin{tabular}[c]{@{}c@{}}0.09\\ (0.1)\end{tabular}  \\ \cline{1-1} \cline{3-12} 
500                  &                      & \multicolumn{1}{c|}{\begin{tabular}[c]{@{}c@{}}0.92\\ (0.1)\end{tabular}}  & \multicolumn{1}{c|}{\begin{tabular}[c]{@{}c@{}}0.02\\ (0.02)\end{tabular}} & 0.7 & \multicolumn{1}{c|}{\begin{tabular}[c]{@{}c@{}}0.94\\ (0.1)\end{tabular}}  & \multicolumn{1}{c|}{\begin{tabular}[c]{@{}c@{}}0.05\\ (0.02)\end{tabular}} & 0.7 & \multicolumn{1}{c|}{\begin{tabular}[c]{@{}c@{}}0.74\\ (0.1)\end{tabular}}  & \begin{tabular}[c]{@{}c@{}}0.15\\ (0.07)\end{tabular} & \multicolumn{1}{c|}{\begin{tabular}[c]{@{}c@{}}0.78\\ (0.06)\end{tabular}} & \begin{tabular}[c]{@{}c@{}}0.06\\ (0.08)\end{tabular} \\ \cline{1-1} \cline{3-12} 
1000                 &                      & \multicolumn{1}{c|}{\begin{tabular}[c]{@{}c@{}}0.96\\ (0.08)\end{tabular}} & \multicolumn{1}{c|}{\begin{tabular}[c]{@{}c@{}}0.06\\ (0.03)\end{tabular}} & 0.6 & \multicolumn{1}{c|}{\begin{tabular}[c]{@{}c@{}}0.94\\ (0.1)\end{tabular}}  & \multicolumn{1}{c|}{\begin{tabular}[c]{@{}c@{}}0.03\\ (0.01)\end{tabular}} & 0.7 & \multicolumn{1}{c|}{\begin{tabular}[c]{@{}c@{}}0.78\\ (0.06)\end{tabular}} & \begin{tabular}[c]{@{}c@{}}0.21\\ (0.13)\end{tabular} & \multicolumn{1}{c|}{\begin{tabular}[c]{@{}c@{}}0.8\\ (0)\end{tabular}}     & \begin{tabular}[c]{@{}c@{}}0.05\\ (0.05)\end{tabular} \\ \hline
\end{tabular}
\captionof{table}{Means of TPR and FPR with corresponding standard deviations given in parenthesis associated to the support recovery of $\pmb{\beta}^\star$ for four methods, different values of $n$, $q$, $\alpha^\star =2$, $p = 100$, and 10 simulations. The column $t$ is the threshold for which the corresponding TPR and FPR are obtained. 
\label{TPF_FPR_table}}
\end{table}

\bibliographystyle{chicago}
\bibliography{biblio}
\end{document}